\documentclass[12pt]{article}
\usepackage{amsmath,amssymb,amsfonts,amsthm,bbm}
\usepackage{epic,eepic,epsfig,longtable}
\usepackage{multirow,verbatim}
\usepackage{array}
\usepackage{graphicx}
\usepackage{floatrow}
\usepackage{subcaption}
\usepackage{paralist}
\usepackage{latexsym}
\usepackage{comment}
\usepackage{epsfig}
\usepackage{setspace}
\usepackage{CJK}
\usepackage{color}
\usepackage{bbm}
\usepackage{rotating}
\usepackage[authoryear, round]{natbib}
\makeatletter
\def\singlespace{\def\baselinestretch{1}\@normalsize}

\makeatletter
\def\singlespace{\def\baselinestretch{1}\@normalsize}

\numberwithin{equation}{section}

\renewcommand{\hat}{\widehat}

\renewcommand{\hat}{\widehat}

\newcommand{\bfm}[1]{\ensuremath{\mathbf{#1}}}

    \def\FF{\mathbb{F}}

   \def\bQ{\bfm Q}

   \def\bY{\bfm Y}

\newcommand{\bfsym}[1]{\ensuremath{\boldsymbol{#1}}}

\def\1{\bfsym{1}}	

 \def\halpha{\hat{\alpha}}              
 \def\hbeta{\hat{\beta}}

 \def\htheta{\hat {\theta}}

 \def\homega {\hat {\omega}}

\def\today{\ifcase\month\or
  January\or February\or March\or April\or May\or June\or
  July\or August\or September\or October\or November\or December\fi
  \space\number\day, \number\year}

\newdimen\biblioindent    \biblioindent=30pt

 at 8truept

\newcommand{\beq}{\begin{equation}}
  \newcommand{\eeq}{\end{equation}}
\newcommand{\beqn}{\begin{eqnarray}}
  \newcommand{\eeqn}{\end{eqnarray}}
\newcommand{\beqnn}{\begin{eqnarray*}}
  \newcommand{\eeqnn}{\end{eqnarray*}}

\allowdisplaybreaks
\usepackage{verbatim}
\def\tilde{\widetilde}

\def\FF{\mathcal{F}}
\def\[{\left [}  \def\]{\right ]} \def\({\left (}  \def\){\right )}
 \def\endpf{$\blacksquare$}
\def\hat{\widehat}

\newtheorem{assumption}{Assumption}
\newtheorem{theorem}{Theorem}

\theoremstyle{definition}

\newtheorem{remark}{Remark}

 \def\I{\mathrm{I}}
 
 \def\I{\mathrm{I}}

\graphicspath{{figures/}{./images}}

\title{
Conditional Quantile Analysis for 
Realized GARCH Models }
\author{Donggyu Kim$^a$, Minseog Oh$^a$, and Yazhen Wang$^b$ \\
$^a$College of Business, \\
Korea Advanced Institute of Science and Technology (KAIST)\\
$^b$Department of Statistics, University of Wisconsin-Madison
}

\begin{document}
\maketitle

\begin{spacing}{1.45}

\begin{abstract}
This paper introduces a novel quantile approach to harness the high-frequency information and improve the daily conditional quantile estimation.
Specifically, we model the conditional standard deviation as a realized GARCH model and employ conditional standard deviation, realized volatility, realized quantile, and absolute overnight return as innovations in  the proposed dynamic quantile models. We devise a two-step estimation procedure to estimate the conditional quantile parameters. The first step applies a quasi-maximum likelihood estimation procedure, with the realized volatility as a proxy for the volatility proxy, to estimate the conditional standard deviation parameters. The second step utilizes a quantile regression estimation procedure with the estimated conditional standard deviation  in the first step.
Asymptotic theory is established for the proposed estimation methods, and a simulation study is conducted to check their finite-sample performance. 
Finally, we apply the proposed methodology to calculate the value at risk (VaR) of 20 individual assets and compare its performance with  existing competitors. 
\end{abstract}

\noindent \textbf{Keywords:}  High-frequency financial data, quasi-maximum likelihood estimation, realized volatility,  risk management, value at risk 


\section{Introduction} \label{SEC-1}

Summary statistics such as realized volatility and quantile play a pivotal role in modern risk management. 
 Generalized autoregressive conditional heteroskedasticity (GARCH) models \citep{bollerslev1986generalized, engle1982autoregressive} are widely used to study low-frequency  volatility dynamics. 
GARCH models adopt squared daily log returns as innovations in the conditional volatilities. When the volatility changes rapidly, 
 it is often difficult to catch up with the change 
 by using only the daily log returns as the innovations \citep{andersen2003modeling}.
  On the other hand, high-frequency financial data are available to 
  construct the so-called realized volatility for estimating daily integrated volatility  \citep{ait2010high,  barndorff2008designing,  fan2018robust,  jacod2009microstructure, shin2021adaptive, xiu2010quasi, zhang2005tale, zhang2006efficient}.
Several conditional volatility models have been developed to draw combined inference based on the low-frequency structure with high-frequency data and enhance volatility estimation and  predication.
 Examples include the heterogeneous auto-regressive (HAR) models \citep{corsi2009simple},  high-frequency based volatility (HEAVY) models \citep{shephard2010realising}, realized GARCH models \citep{hansen2012realized}, GARCH-It\^o models \citep{kim2016statistical, kim2019factor, kim2016unified, song2021volatility}, and overnight GARCH-It\^o models \citep{kim2021overnight}.
Their empirical studies show that incorporating the realized volatility as the new innovation into the models improves volatility modeling and provides better explanation of the volatility dynamics in the financial market.

Conditional quantile is an essential ingredient along with the conditional volatility for the study of risk measures such as conditional value at risk (VaR). 
We often assume that the log return follows a conditionally normal distribution or some known heavy-tailed symmetric distribution such as the t-distribution.  
Under the distribution assumption, the conditional quantile estimation is reduced to the conditional volatility estimation. 
However, it is often empirically observed that the log return has negative skewness and excess kurtosis, which violates the distribution assumption.
Thus, to obtain more robust estimation of conditional quantile, quantile regression for time series has been introduced. 
 \citet{koenker1996conditional} extended quantile regression to linear ARCH models for estimating conditional quantiles of log returns. 
\citet{engle2004caviar}  suggested a nonlinear dynamic quantile model with an AR structure for conditional quantilesl.   
\citet{xiao2009conditional}  further proposed the two-step conditional quantile estimation for GARCH models.  
These quantile regression based models are based on low-frequency data to capture the market dynamics.  
On the other hand, \citet{vzikevs2015semi} harnessed the quantile autoregressions with realized volatility  and found the benefits of incorporating high-frequency information.   
See also \citet{giot2004modelling, louzis2014realized}.
To account for the autoregressive (AR) structure  of the conditional quantile, we often employ CAViaR model \citep{engle2004caviar} with the realized volatility \citep{vzikevs2015semi}.
This AR structure helps to obtain the parsimonious property, but in terms of estimating the model parameter, the realized CAViaR \citep{vzikevs2015semi} uses computationally intensive estimation methods.  
Thus, in this paper, we link the realized GARCH models with the realized  CAViaR structure and propose a two-step estimation procedure to reduce the estimation complexity.


Specifically, we assume that the conditional standard deviation follows a realized GARCH model with the square root of open-to-close integrated volatility and absolute of the overnight return as the innovations. The modeling allows to derive an AR structure such as the quantile regression models in \citet{engle2004caviar} and \citet{xiao2009conditional}.
To reduce the complexity of estimating model parameters, 
we rewrite the model as some quantile regression of the previous conditional standard deviation, square root of open-to-close integrated volatility, and absolute of the overnight return.
We call this model the realized GARCH quantile regression  model.
On the other hand,  we impose some self-similarity condition on the high-frequency data in order to utilize the realized quantile information. 
The proposed quantile regression model has the open-to-close realized quantile \citep{dimitriadis2021realized} as the innovation instead of the realized volatility. 
We call this  model the realized quantile and realized GARCH quantile regression (real-realized GARCH quantile regression) model.
To estimate the model parameter, we suggest a two-step estimation procedure under a location-scale assumption, which helps to simplify estimation procedures. 
In the first step, to estimate the conditional standard deviation, we harness the high-frequency information.
We employ a quasi maximum likelihood estimation procedure with the realized volatility as the open-to-close conditional volatility proxy and squared overnight return as the close-to-open conditional volatility proxy. 
In the second step, we apply the quantile regression estimation procedure with the estimated conditional standard deviation in the first step. 
The proposed two-step estimation procedure is relatively easy to implement, and as shown in \citet{kim2016unified}, employing the high-frequency information helps to obtain more accurate parameter estimator. 
We show the asymptotic theory for the two-step
estimator in the presence of the measurement (discretization) error associated
with realized volatility.

The rest of the paper is organized as follows. Section \ref{SEC-2} proposes the dynamic quantile  
models. Section \ref{SEC-3}  proposes the two-step estimation procedure and derives its asymptotic properties.
In Section \ref{SEC-4},  a simulation study is conducted to check the finite sample performance of the proposed estimator.
 Section \ref{SEC-5} applies the proposed method 
 to the conditional VaR predication for individual assets. 
 All the technical proofs are collected in Section \ref{SEC-Proof}.


\setcounter{equation}{0}
\section{Dynamic realized quantile regression models} \label{SEC-2}

\subsection{Realized GARCH models}
In the high-frequency finance, we often assume that the log stock price satisfies the following diffusion process,
\begin{equation*}
dX_t=   \mu_t dt+  \sigma_{t}  dB_{t},  
\end{equation*}
where $B_t$ is a standard Brownian motion, and $\mu_t$ and  $\sigma_t$ are the drift and instantaneous volatility processes, respectively, that is adapted to $\FF_t$.
With the high-frequency financial data, there are several well-performed realized volatility estimators, for example,   multi-scale realized volatility (MSRV) \citep{zhang2006efficient, zhang2011estimating},  
pre-averaging realized volatility (PRV) \citep{christensen2010pre, jacod2009microstructure}, kernel realized volatility (KRV) \citep{barndorff2008designing, barndorff2011multivariate}, 
quasi-maximum likelihood estimator (QMLE) \citep{ait2010high, xiu2010quasi},  local method of moments \citep{bibinger2014estimating},  and robust pre-averaging realized volatility \citep{fan2018robust, shin2021adaptive}.
With these realized volatility estimators as the new innovation in the GARCH model,  \citet{hansen2012realized} introduced the realized GARCH models, and 
\citet{kim2016unified} proposed a unified GARCH-It\^o model.
 Recently, \citet{kim2021overnight} further employed the squared overnight log return  as the overnight risk innovation.
Their empirical studies show that incorporating the realized  volatility and overnight log return helps to explain market dynamics. 
In the similar spirit,  we assume that the daily close-to-close log-return $Y_n$ has the following realized GARCH models 
\begin{eqnarray}\label{Realized GARCH}
	&&Y_n=    h_n (\theta) \epsilon_n, \cr
	&&h_n (\theta) = \omega+\gamma h_{n-1} (\theta) + \alpha \sqrt{IV_{n-1}} +\beta \sqrt{OV_{n-1}},  \cr
	&& \int_{n-1}^n \sigma_t^2 dt = h_n(\theta) ^2 + D_n, 
\end{eqnarray}
where the open-to-close integrated volatility $IV_{n-1} = \int_{n-1-\lambda}^{n-1} \sigma_t^2 dt$, overnight volatility $OV_{n-1} = (X_{n-1-\lambda} - X_{n-2})^2 $, $\lambda$ is the open-to-close trading hours, $D_n$ is a martingale difference, and for given $\FF_{n-1}$,  $\epsilon_{n}$'s are i.i.d. random variables with mean zero.
The conditional standard deviation $h_n(\theta)$ has some overnight GARCH-It\^o (OGI)  \citep{kim2021overnight} form with the two innovations terms; the square root of the open-to-close integrated volatility and squared overnight return. 
That is, the market volatility dynamics are explained by the open-to-close integrated volatility and squared close-to-open log returns,
which represent volatilities for the open-to-close and close-to-open periods, respectively.

\begin{remark}\label{Remark1} 
As studied in \citet{kim2021overnight} and \citet{song2021volatility}, we can find some It\^o diffusion process which satisfies the  realized GARCH model in  \eqref{Realized GARCH} and the integrated volatility has the following relationship 
$$
\int_{n-1}^n \sigma_t^2 dt = h_n(\theta) ^2 + D_n.
$$
With this relationship, we can employ the non-parametric realized volatility estimators to estimate the GARCH parameter and study asymptotic properties.
One possible instantaneous volatility process is the step function. 
Specifically, the instantaneous volatility satisfies, for $t \in [n-1, n)$,
\begin{equation*}
\sigma_t^2 = h_{n} (\theta) ^2 + D_n.
\end{equation*}
There may exist more realistic diffusion processes as in  \citet{kim2021overnight} and \citet{song2021volatility}.
However, the main purpose of this paper is to develop some low-frequency time series model which can account for conditional quantile dynamics of log returns. 
Thus, we leave developing diffusion processes for future study. 
\end{remark}

Given the current information $\FF_{n-1}$, the one-day ahead conditional quantile value of $Y_n $ is 
\begin{equation*}
 	Q_{\tau, n} ( \theta) =  h_{n} (\theta)  q_{\tau}, 
\end{equation*}
 where $q_{\tau}$ is the $\tau$-quantile value of $\epsilon_n$. 
We can rewrite the above conditional quantile value as follows:
\begin{equation*}
	Q_{\tau, n}  (\theta) = \omega_\tau + \gamma Q_{\tau, n-1} (\theta) + \alpha_\tau \sqrt{IV_{n-1}} +\beta_{\tau} \sqrt{OV_{n-1}},
\end{equation*} 
where $\omega_{\tau}= \omega q_{\tau}$,  $\alpha_{\tau}= \alpha q_{\tau}$, and $\beta_{\tau}= \beta q_{\tau}$. 
The conditional quantile value has the auto-regression form with the square root of the open-to-close integrated volatility and squared overnight return.
That is, it has the conditional autoregressive VaR (CAViaR) structure proposed by \citet{engle2004caviar} and \citet{vzikevs2015semi}. 
Thus, the proposed model can be considered as the special case of CAViaR models.
On the other hand, under the realized GARCH volatility structure \eqref{Realized GARCH}, the conditional quantile can be rewritten as follows:
\begin{equation}\label{CQ}
	Q_{\tau, n}  (\theta) = \omega_{\tau} + \gamma_{\tau} h_{n-1} (\theta) + \alpha_{\tau} \sqrt{IV_{n-1}} + \beta_{\tau} \sqrt{OV_{n-1}} ,
\end{equation}
where   $\gamma_{\tau}= \gamma q_{\tau}$.
The quantile regression has three explanatory variables; the previous conditional standard deviation, square root of the open-to-close integrated volatility, and absolute value of overnight return. 
The conditional standard deviation $h_{i-1}(\theta)$ has the realized GARCH form, which helps the quantile regression model to parsimoniously capture the persistent influence of long-past shocks. 
Unlike the CAViaR form, the regression form in \eqref{CQ} is a linear form when we consider $h_{i-1}(\theta)$ as one variable. 
This property reduces the model complexity and makes it easy to make inferences. 
We call this model the realized GARCH quantile regression. 
This structure is similar to the GARCH-based quantile regression model proposed by \citet{xiao2009conditional}. 
However, while they employ  only  the low-frequency information, in this paper, we  study how to  incorporate the high-frequency data in both quantile regression modeling and parameter inferences.

  \subsection{Realized quantile}

 Recently, under the self-similarity condition, \citet{dimitriadis2021realized} suggested the realized quantile. 
For example, for some $H \in (0,1)$,  the stochastic process $X_t$ satisfies
\begin{equation}\label{self}
	X_{t+ c\Delta}- X_t  \overset{d}{=} c^H \[ X_{t+ \Delta} - X_t \],
\end{equation}
where $\overset{d}{=}$ denotes equality in distribution. 
If  the instantaneous process is a step function as discussed in Remark \ref{Remark1} and $\mu_t=0$,  the self-similarity condition is satisfied and $H =0.5$.  
The realized quantile satisfies the following scale relationship
\begin{equation*}
	RQ_{\tau } (X_n - X_{n-\lambda})  =\frac{1}{\sqrt{ \Delta}} RQ_{\tau } (X_{t_{n,i} } - X_{t_{n, i-1} }),
\end{equation*}
where $RQ_{\tau } (X)$ is the quantile of $X$ at the probability $\tau$ and $t_{n,i} = n-1 + \Delta  i$ for $i=0,1,\ldots, \Delta^{-1}$. 
Then, using the high-frequency return data, we can estimate the realized quantile of the open-to-close return, $X_n - X_{n-\lambda}= \int_{n-\lambda} ^n \sigma_t dB_t$. 
Specifically, we calculate the sample quantile using the high-frequency return data, $X_{t_{n,i} } - X_{t_{n, i-1} }$, then by multiplying $\frac{1}{\sqrt{\Delta}}$, we can estimate the realized quantile  of the open-to-close return. 
The realized quantile harnesses the high-frequency information in estimating the quantile, thus it has the quanitle dynamics information directly. 
Furthermore, the conditional realized quantile of the open-to-close return, $X_n - X_{n-\lambda}= \int_{n-\lambda} ^n \sigma_t dB_t$,  is 
\begin{equation*}
	RQ_{\tau } (X_n - X_{n-\lambda})   = z_\tau \sqrt{ \int_{n-\lambda} ^n \sigma_t^2 d t}  ,
\end{equation*}
where $z_{\tau}$ is the $\tau$-quantile of $\frac{1}{\sqrt{\lambda}} \int_{n-\lambda} ^n  dB_t$.
Then the realized GARCH quanitle regression model \eqref{CQ} becomes
\begin{equation}\label{CQ-2}
	Q_{\tau, n}  (\theta) = \omega_{\tau} + \gamma_{\tau} h_{n-1} (\theta) + \alpha_{\tau} ^{\prime}  RQ_{\tau, n-1}  + \beta_{\tau} \sqrt{OV_{n-1}} ,
\end{equation}
where $RQ_{\tau, n}= RQ_{\tau } (X_n - X_{n-\lambda}) $ and  $\alpha_{\tau} ^\prime= \alpha q_{\tau} /z_{\tau} $.
Unlike \eqref{CQ}, the dynamic quantile regression model has the realized quantile instead of the integrated volatility  as the explanatory variable. 
We call this model the realized quantile and realized GARCH quantile regression (real-realized GARCH quantile regression) models.
Since the realized quantile contains the quantile information directly, the real-realized GARCH quantile regression may capture the quantile dynamics well. 
Unfortunately, to hold the above relationship, we need the step function condition of volatility  over each day, which is often violated in the real data analysis. 
However, as the new innovation information, the realized quantile may be helpful to explain the quantile dynamics.

\begin{remark}
The realized GARCH quantile regression and real-realized GARCH quantile regression models are based on the realized GARCH structure in order to capture the volatility dynamics. 
Thus, they can account for the volatility clustering. 
The difference between them is the source of explanatory variables for the quantile dynamics.
Specifically, the real-realized GARCH quantile regression incorporates the realized quantile, while the realized GARCH quantile regression employs the realized volatility. 
If the self-similarity condition is satisfied, since the real-realized GARCH quantile regression contains the quantile information directly, it may be able to capture the quantile dynamics well. 
However, it is hard to satisfy the self-similarity condition. 
In contrast, the realized GARCH quantile regression does not need the self-similarity condition, and, so, it is relatively robust to the volatility structure.
However, the realized GARCH quantile regression model relies on the volatility dynamic structure, thus, it is hard to explain other source of the quantile dynamics. 
\end{remark}

\section{Estimation procedure} \label{SEC-3}
%
%

 \subsection{Two-step estimation procedure for the realized GARCH quantile regression model} \label{SEC-3.1}
 
 We recall that the conditional quantile in \eqref{CQ} can be considered as the linear equation with explanatory variables such as square root of integrated volatility, absolute value of overnight return, and conditional standard deviation as follows:
 \begin{equation*}
	Q_{\tau, n}  (\theta) = \omega_{\tau} + \gamma_{\tau} h_{n-1} (\theta) + \alpha_{\tau} \sqrt{IV_{n-1}} + \beta_{\tau} \sqrt{OV_{n-1}}.
\end{equation*}
That is, the conditional quantile is explained by $h_{n-1} (\theta)$,  $\sqrt{IV_{n-1}}$, and $\sqrt{OV_{n-1}}$.
In the first stage, we estimate these explanatory variables. 
The open-to-close integrated volatility $IV_{n}$ is not observed, so  we need to estimate it. 
For example, we calculate the realized volatility (RV) estimator as follows:
$$
RV_n = \sum_{i=1}^m (X_{t_{n,i}} - X_{t_{n,i-1}})^2,
$$
where $m$ is the number of high-frequency observations for the open-to-close period.
Then the realized volatility estimator converges to the integrated volatility with the convergence rate $m^{1/2}$.

%

 To evaluate the conditional standard deviation $h_{n}(\theta)$, we employ the high-frequency observations. 
Specifically, under the realized GARCH model \eqref{Realized GARCH}, the integrated volatility is a good proxy of the conditional standard deviation as follows:
$$
\int_{n-1}^n \sigma_t^2 dt = h_n(\theta) ^2 + D_n.
$$
For the open-to-close period, we can use the non-parametric realized volatility $RV_n$ as the estimator of the open-to-close integrated volatility. 
However, for the close-to-open period, we cannot observe the high-frequency data, thus we use the squared overnight return as the proxy. 
By It\^o's lemma, we have
\begin{equation*}
	OV_n = \int_{n-1}^{n-\lambda} \sigma_t^2 dt + 2 \int_{n-1}^{n-\lambda} (X_t -X_{n-1}) dX_t \text{ a.s.} 
\end{equation*}
Therefore, we obtain 
\begin{equation*}
	IV_n+OV_n = \int_{n-1}^{n} \sigma_t^2 dt + 2 \int_{n-1}^{n-\lambda} (X_t -X_{n-1}) dX_t= h_n(\theta) ^2 +D_n ^L, 
\end{equation*}
where $D_n^L = D_n + 2 \int_{n-1}^{n-\lambda} (X_t -X_{n-1}) dX_t$ is a martingale difference.
Based on this relationship, we employ the realized volatility and squared overnight return as the proxy of the quasi maximum likelihood estimation to get an estimator of $\theta$ as follows: 
 \begin{equation*}
	\hat{\theta} = \arg\max _{\theta \in \Theta} \hat{L}_{n,m} (\theta)  ,
\end{equation*}
where 
\begin{eqnarray*}
&&\hat{L}_{n,m} (\theta) =- \frac{1}{n} \sum_{i=1}^n \log \hat{h}_i^2  (\theta) +\frac{RV_i +  OV _i }{\hat{h}_i^2 (\theta)  }, \cr
 &&\hat{h}_i (\theta) =     \omega  + \gamma \hat{h}_{i-1}( \theta) + \alpha  \sqrt{RV_{i-1}} + \beta \sqrt{ OV _{i-1}}.
\end{eqnarray*}
That is, the non-parametric volatility estimator $RV_i +  OV _i$ is employed as the proxy of conditional GARCH volatility, and in the Gaussian quasi likelihood sense,  we use the non-parametric volatility estimator instead of the squared daily log return. 
As shown in \citet{kim2016unified}, adopting realized volatility as the proxy in the quasi maximum likelihood estimation improves the accuracy of estimating parameters comparing with the QMLE procedure with the squared log return as the proxy. 
We also enjoy the same benefit by harnessing the high-frequency data. 
Then, with the QMLE estimator $\hat{\theta}$, we estimate the conditional standard deviation as follows:
$$
\hat{h}_{n} (\hat{\theta})= \hat{\omega}  + \hat{\gamma} \hat{h}_{n-1}( \hat{\theta}) + \hat{\alpha}  \sqrt{RV_{n-1}} + \hat{\beta} \sqrt{ OV _{n-1}}.
$$

 In the second stage, with the estimated explanatory variables in the first stage, we  apply the quantile regression  to estimating the true quantile parameters, $\theta_{\tau 0} = (\omega_{\tau 0}, \beta_{\tau 0}, \gamma_{\tau 0} )$, as follows:
\begin{equation*}
\hat{\theta}_{\tau}^{RG} = \arg \min _{\theta_{\tau}} \sum_{i=2}^n  \rho_{\tau} \( Y_i   -   \omega_{\tau} -  \gamma_{\tau} \hat{h} _{i-1} (\hat{\theta})  - \alpha_{\tau} \sqrt{ RV_{i-1}}  -\beta_{\tau} \sqrt{ OV _{i-1}}   \),
\end{equation*}  
where $\hat{\theta}$ is the QMLE result in the first stage, $\rho _{\tau} (x) =  x ( \tau - \I (x <0)),$ and $I(\cdot)$ is an indicator function.

To study the first step estimation procedure, we need the following assumptions.
\begin{assumption} \label{assumption-1}
	~
	\begin{enumerate}
		\item[(a)] Let
		\begin{equation*}
		\Theta = \lbrace  (\omega , \gamma,  \alpha, \beta): \omega_l < \omega < \omega_u, \gamma_l < \gamma < \gamma_u, \alpha_l < \alpha < \alpha_u, \beta_l <\beta < \beta_u, \gamma+\alpha+ \beta<1   \rbrace, 
		\end{equation*}
		where $\omega_ l, \omega_u,  \gamma_l, \gamma_u, \alpha_l, \alpha_u, \beta_l, \beta_u$ are known positive constants.
		
		\item[(b)] One of the following conditions is satisfied.
		\begin{itemize}
			\item[(b1)] There exists a positive constant $\delta$ such that $E \left[ \left( \frac{Y_{i}^{2}}{h_{i}^2(\theta_0)} \right) ^{2+\delta} \right] \leq C$ for any $i \in \mathbb{N}$.
			
			\item[(b2)] $\frac{E[Y_i^4 | \mathbf{\mathcal{F}}_{i-1}]}{h^4_i(\theta_0)} \leq C$ a.s. for any $i \in \mathbb{N}$.
		\end{itemize}

		\item[(c)] We have $\underset{t \in \mathbb{R_+}}{\max} \,  E \left\lbrace \sigma^4_t \right\rbrace < \infty $  .
		\item[(d)] $\sup\limits_{i\in \mathbb{N}} E \[ (RV_{i}- IV_i )^2 \] \leq C   m^{-1}$.
		
		\item[(e)] For any $i\in \mathbb{N}$, $E\left[RV_{i}|\mathcal{F}_{i-1}\right]\leq C \, E \left[ IV_i | \mathcal{F}_{i-1} \right]+C$ a.s.
		
		\item[(f)] $\left(D_i^L, IV_i, OV_i, Y^2_i \right)$ is a non-degenerating strictly  stationary ergodic process.
	\end{enumerate}
\end{assumption}

\begin{remark}
We study the second moment related parameters, thus, the fourth moment conditions such as Assumption \ref{assumption-1}(b)--(c) are required.
 Under some mild moment condition, we can show   Assumption \ref{assumption-1}(d)--(e). 
 For example, under the finite fourth moment condition, \citet{kim2016asymptotic} showed that the realized volatility estimators satisfy  Assumption \ref{assumption-1}(d).
 The moment condition of the realized volatility estimator is often required to investigate the double-asymptotics of letting both $n$ and $m$ go to infinity (see \citet{corradi2011predictive,corradi2012international}).
 Finally, to establish the asymptotic normality, we need Assumption \ref{assumption-1}(f).
\end{remark}

The below theorem establishes asymptotic properties of the QMLE method in the first step. 

\begin{theorem} \label{Thm-1}
Under the model \eqref{Realized GARCH}, Assumption \ref{assumption-1}(a)--(e) are met. 
Then we have
\begin{equation}\label{Thm1-result1}
	\sqrt{n} ( \hat{\theta} - \theta_0 )=  \frac{B_1 ^{-1}}{\sqrt{n}}  \sum_{i=1}^n \frac{D_i ^L}{h_i^4 (\theta_0) } \frac{\partial h_i^2 (\theta)}{ \partial \theta} \Big|_{\theta= \theta_0}+O_p(n^{1/2} m^{-1/2} ) +o_p(1),
\end{equation}
where $\theta_0 =(\omega_0, \beta_0, \gamma_0)$ is the true parameter, and
$$
B _1 = E \[ \frac{1}{h_1 ^4(\theta_0) }  \frac{\partial h_1^2 (\theta)}{ \partial \theta}  \frac{\partial h_1^2 (\theta)}{ \partial \theta ^{\top}} \Big|_{\theta= \theta_0} \].
$$
Furthermore, Assumption \ref{assumption-1} is satisfied and suppose that $n  m^{-1}  \to 0$.
Then we have
\begin{equation}\label{Thm1-result2}
	\sqrt{n} ( \hat{\theta} - \theta_0 ) \overset{d}{\to} N ( 0,  B_1^{-1}B_2 B_1^{-1}) ,
\end{equation}
where 
$$
B _2 =   E \[ \frac{(D_1^L )^2 }{h_1 ^8(\theta_0) }  \frac{\partial h_1^2 (\theta)}{ \partial \theta}  \frac{\partial h_1^2 (\theta)}{ \partial \theta ^{\top}} \Big|_{\theta= \theta_0} \].
$$
\end{theorem}

\begin{remark}
Theorem \ref{Thm-1} shows that the convergence rate is $n^{-1/2}$ and $m^{-1/2}$. 
The $n^{-1/2}$ term is the usual optimal convergence rate, and the $m^{-1/2}$ term is coming from estimating the realized volatility. 
To obtain the asymptotic normality, we need to mitigate the noise coming from the high-frequency observations.
Thus, we additionally need  the condition $n  m^{-1}  \to 0$, which makes the noises from the realized volatility estimator negligible.  
Under this condition, we derive the asymptotic normality.
\end{remark}

In the first step, we adopt the high-frequency data, which  reduces the asymptotic variance.
For example, by the It\^o's lemma,  the squared log return is
\begin{equation*}
 Y_n^2 = \int_{n-1}^{n} \sigma_t^2 dt + 2 \int_{n-1}^{n} (X_t -X_n) dX_t = h_n (\theta)^2 +D_n+2 \int_{n-1}^{n} (X_t -X_n) dX_t .  
\end{equation*}
When employing the squared daily log return as the proxy in the QMLE procedure, we have the martingale difference term $D_n+2 \int_{n-1}^{n} (X_t -X_n) dX_t$.
The term $D_n+2 \int_{n-1}^{n} (X_t -X_n) dX_t$ has higher variance than $D_n^L$, which increases the asymptotic variance term $B_2$.  
That is, as harnessing the high-frequency information in the first step, we are able to increase the accuracy of the estimation procedure.

To investigate the two-step estimation procedure, we need the following additional technical conditions.

\begin{assumption} \label{assumption-2}
	~
	\begin{enumerate}
		\item[(a)]  Denote the conditional distribution function $P \( Y_n < \cdot | \FF_{n-1} \)$ by $F_{Y| X} (\cdot) $. 
		Its derivative $f_{Y|X} (\cdot) $ is continuously differentiable, and $0<f_{Y|X} (\cdot)  < \infty$ on its support.
		
		\item[(b)]   The distribution function of $\epsilon_i$, $F_{\epsilon} (\cdot)$, has a continuous density $f_{\epsilon} (\cdot)$ with $0< f_{\epsilon} (F^{-1}_{\epsilon} (\tau) ) <\infty$.  
		
		\item[(c)] There exists positive constant $\delta_1>0$ and $\delta_2>0$ such that 
		$$
		P(\max_{1 \leq t\leq n} Y_t^2 > n^{\delta_1}) \leq \exp (- n^{\delta_2}).
		$$
		
	   \item [(d)]   $E \[ \frac{ A_i(\theta_0) A_i (\theta_0) ^{\top}  }{h _{i} (\theta_0) } \]$ has a full rank, where  $A_i (\theta) = ( 1, h _{i-1} (\theta),  \sqrt{ IV_{i-1}} , \sqrt{OV_{i-1}}) ^{\top}$.
	\end{enumerate}
\end{assumption}

\begin{remark}
Assumption \ref{assumption-2} is usually required to analyze the quantile regression and two-step estimation procedure (see \citet{xiao2009conditional}).  
\end{remark}

The theorem below establishes the asymptotic properties of the two-step estimator $\hat{\theta}_\tau$.

\begin{theorem}\label{Thm-2}
Under the models \eqref{Realized GARCH} and \eqref{CQ}, assumptions in Theorem \ref{Thm-1} and Assumption \ref{assumption-2} are met.
Then we have
\begin{eqnarray}  \label{Thm-2-result1}
	\sqrt{n} ( \hat{\theta}_{\tau}^{RG}- \theta_{\tau0} )&=& \frac{1}{f_{\epsilon} ( F_{\epsilon} ^{-1}  (\tau) ) }  \Gamma_1 ^{-1} \frac{1}{\sqrt{n}}  \sum_{i=2}^n \left  \{  \tau -  \I (Y_i   < \theta_{\tau 0}^{\top} A_i (\theta_0))    \right \} A_i (\theta_0)  \cr
	&&  -\Gamma_1 ^{-1} \Gamma_2 \sqrt{n} (\hat{\theta}  - \theta_0 ) + o_p(1) ,
\end{eqnarray}
where $\Gamma_1 = E \[ \frac{ A_i(\theta_0) A_i (\theta_0) ^{\top}  }{h _{i} (\theta_0) } \]$,  $\Gamma_2= E\[  \frac{A_i (\theta_0)}{h_i(\theta_0)} \frac{\partial \theta_{\tau 0} ^{\top} A_{i} (\theta) }{\partial \theta ^{\top}} \Big| _{\theta = \theta_0}    \] $.
The limiting distribution is 
\begin{equation*}
	\sqrt{n} ( \hat{\theta}_{\tau}^{RG}- \theta_{\tau0} ) \overset{d} {\to} N(0,  \Gamma_1 ^{-1} M  \Gamma_1 ^{-1} ),
\end{equation*}
where 
\begin{eqnarray*}
M &=&   \frac{\tau (1-\tau) }{f_{\epsilon} ^2( F_{\epsilon} ^{-1}  (\tau) ) }  E \[  A_i(\theta_0) A_i (\theta_0) ^{\top}   \] + \Gamma_2 B_1^{-1}B_2 B_1^{-1} \Gamma_2^{\top} \cr
&&+  \frac{1}{f_{\epsilon} ( F_{\epsilon} ^{-1}  (\tau) ) }  E \Bigg [  \left  \{  \tau -  \I (Y_i   < \theta_{\tau 0}^{\top} A_i (\theta_0))    \right \} \frac{D_i ^L}{h_i^4 (\theta_0) }  \cr
&& \qquad \qquad \qquad \quad    \times \Bigg \{ A_i (\theta_0)   \frac{\partial h_i^2 (\theta)}{ \partial \theta ^{\top} } \Big|_{\theta= \theta_0}  B_1^{-1} \Gamma_2  ^{\top}    +  \Gamma_2    B_1^{-1}  \frac{\partial h_i^2 (\theta)}{ \partial \theta } \Big|_{\theta= \theta_0}  A_i (\theta_0)  ^{\top} \Bigg \}  \Bigg ] .
\end{eqnarray*}
\end{theorem}
\begin{remark}
Theorem \ref{Thm-2} shows the asymptotic normality of the two-step estimator $\hat{\theta}_{\tau}^{RG}$.
Since it utilizes the QMLE estimator in the first step, the effect of $\hat{\theta}$ remains as in \eqref{Thm-2-result1}.
Specifically, the remaining term is $\Gamma_1 ^{-1} \Gamma_2 \sqrt{n} (\hat{\theta}  - \theta_0 ) $.
As we discussed, thanks to using the high-frequency observations, we can reduce the asymptotic variance of $\sqrt{n} (\hat{\theta}  - \theta_0 )$, thus the two-step estimator $\hat{\theta}_\tau$ also has smaller asymptotic variance comparing with the estimation procedure which harnesses only the low-frequency information.
\end{remark}

 \subsection{Two-step estimation procedure for the real-realized quantile regression model}

 For the  real-realized quantile regression, we employ the first step estimator in Section \ref{SEC-3.1}. 
 To evaluate the real-realized quantile regression, we need to estimate the realized quantile $RG_{\tau,n}$. 
 For example, we calculate the the quantile of the high-frequency return as follows:
 \begin{equation*}
 	\hat{Q}_{\tau,n} = \arg \min_b \sum_{i=2}^m  \rho_{\tau} \( X_{t_{n,i}} - X_{t_{n,i-1}}   - b    \). 
 \end{equation*}
 Then, under the self-similarity condition, to obtain the quantile for the open-to-close return, we  scale up as follows:
 \begin{equation*}
 	\hat{RQ}_{\tau ,i-1}  = \frac{1}{\sqrt{\Delta}}\hat{Q}_{\tau,n}.  
 \end{equation*}
 With the QMLE estimator in the first step and the realized quantile estimator, we estimate the real-realized quantile regression parameter as follows:
 	\begin{equation*}
\hat{\theta}_{\tau}^{RR} = \arg \min _{\theta_{\tau}} \sum_{i=2}^n  \rho_{\tau} \( Y_i   -   \omega_{\tau} -  \gamma_{\tau} \hat{h} _{i-1} (\hat{\theta})  - \alpha_{\tau} \hat{RQ}_{\tau ,i-1}  -\beta_{\tau} \sqrt{ OV _{i-1}}   \).
\end{equation*}  
To estimate its asymptotic properties, we need the following additional conditions.

 \begin{assumption} \label{assumption-3}
 ~
 \begin{itemize}
 \item [(a)]
 We have $\sup_{i} |\hat{RQ}_{\tau,i} - RQ_{\tau, i}| = O_p(m^{-1/2})$. 
 \item [(b)]   $E \[ \frac{ A_i^{RR} (\theta_0) A_i^{RR}(\theta_0) ^{\top}  }{h _{i} (\theta_0) } \]$ has a full rank, where  $A_i^{RR} (\theta) = ( 1, h _{i-1} (\theta),  RQ_{\tau, i-1}, \sqrt{OV_{i-1}}) ^{\top}$.
 \item[(c)]  $\left(D_i^L, IV_i, OV_i, Y^2_i , RQ_{\tau,i} \right)$ is a  non-degenerating strictly  stationary ergodic process.
 \end{itemize}
\end{assumption}

\begin{remark}
The consistency assumption Assumption \ref{assumption-3}(a) can be obtained under some self-similarity condition and some stationary condition.
For example, the self-similarity condition \eqref{self} with $H=0.5$ is satisfied,  the log-returns are $\alpha$-mixing, and their distribution is absolutely
continuous with strictly positive and continuous density.
Details can be found in Theorem 2.1 \citep{dimitriadis2021realized}.
 Assumption \ref{assumption-3}(c) is required to establish the asymptotic normality. 
 \end{remark}

 The following theorem studies the asymptotic properties of $\hat{\theta}_{\tau}^{RR}$. 
 \begin{theorem}\label{Thm-3}
 Under the models \eqref{Realized GARCH} and \eqref{CQ-2}, assumptions in Theorems \ref{Thm-1}--\ref{Thm-2} and Assumption \ref{assumption-3} are met. 
 Then we have
\begin{eqnarray*}   
	\sqrt{n} ( \hat{\theta}_{\tau}^{RR}- \theta_{\tau0} )&=& \frac{1}{f_{\epsilon} ( F_{\epsilon} ^{-1}  (\tau) ) }  \Gamma_1 ^{RR-1} \frac{1}{\sqrt{n}}  \sum_{i=2}^n \left  \{  \tau -  \I (Y_i   < \theta_{\tau 0}^{\top} A_i^{RR} (\theta_0))    \right \} A_i^{RR} (\theta_0)  \cr
	&&  -\Gamma_1 ^{RR-1} \Gamma_2 ^{RR} \sqrt{n} (\hat{\theta}  - \theta_0 ) + o_p(1) ,
\end{eqnarray*}
where $\Gamma_1 ^{RR}= E \[ \frac{ A_i^{RR}(\theta_0) A_i^{RR} (\theta_0) ^{\top}  }{h _{i} (\theta_0) } \]$,  $\Gamma_2^{RR}= E\[  \frac{A_i ^{RR}(\theta_0)}{h_i(\theta_0)} \frac{\partial \theta_{\tau 0} ^{\top} A_{i}^{RR} (\theta) }{\partial \theta ^{\top}} \Big| _{\theta = \theta_0}    \] $.
The limiting distribution is 
\begin{equation*}
	\sqrt{n} ( \hat{\theta}_{\tau}^{RR}- \theta_{\tau0} ) \overset{d} {\to} N(0,  \Gamma_1 ^{RR-1} M^{RR}  \Gamma_1 ^{RR-1} ),
\end{equation*}
where 
\begin{eqnarray*}
M^{RR} &=&   \frac{\tau (1-\tau) }{f_{\epsilon} ^2( F_{\epsilon} ^{-1}  (\tau) ) }  E \[  A_i^{RR}(\theta_0) A_i^{RR} (\theta_0) ^{\top}   \] + \Gamma_2^{RR} B_1^{-1}B_2 B_1^{-1} \Gamma_2^{RR \top} \cr
&&+  \frac{1}{f_{\epsilon} ( F_{\epsilon} ^{-1}  (\tau) ) }  E \Bigg [  \left  \{  \tau -  \I (Y_i   < \theta_{\tau 0}^{\top} A_i ^{RR}(\theta_0))    \right \} \frac{D_i ^L}{h_i^4 (\theta_0) }  \cr
&& \qquad \qquad \qquad \quad    \times \Bigg \{ A_i (\theta_0)   \frac{\partial h_i^2 (\theta)}{ \partial \theta ^{\top} } \Big|_{\theta= \theta_0}  B_1^{-1} \Gamma_2  ^{RR \top}    +  \Gamma_2^{RR}    B_1^{-1}  \frac{\partial h_i^2 (\theta)}{ \partial \theta } \Big|_{\theta= \theta_0}  A_i^{RR} (\theta_0)  ^{\top} \Bigg \}  \Bigg ] .
\end{eqnarray*}
\end{theorem}

 \begin{remark}
 Similar to Theorem \ref{Thm-2}, Theorem \ref{Thm-3} obtains the asymptotic normality of    the two-step estimator $\hat{\theta}_{\tau}^{RR}$.
\end{remark}

\section{A simulation study} \label{SEC-4}
We conducted simulations to check finite sample performances of the proposed estimation procedures and compare them with existing methods.
We generated the log-prices $X_{t_{i,j}}$ for $n$ days with frequency $1/m$ for each day and let $t_{i,j} = i-\lambda+\frac{\lambda}{m}j$, where $i=1,\ldots,n$, $j=0,\ldots,m$, and the open-to-close trading hours $\lambda=6.5/24$.
The underline true log stock prices follow the diffusion process 
\begin{eqnarray*}
	&&dX_t = \sigma_t dB_t, \cr
	&&\sigma_t^2=\begin{cases}
\frac{w}{\lambda} h^2_i\left(\theta\right) (1 + d_i )  & \text{ if } t \in [[t] +1 -\lambda ,  [t]+1)  \\ 
 \frac{1-w}{1-\lambda}h^2_i\left(\theta\right) (1 + d_i ) & \text{ if } t \in [  [t], [t] +1 -\lambda ),
\end{cases}
\end{eqnarray*}
where $h_i(\theta) = \omega + \gamma h_{i-1}(\theta) + \alpha\sqrt{IV_{i-1}} + \beta\sqrt{OV_{i-1}}$,  and $d_i+0.1$ is generated by noncentral chi-squared distribution with mean 0.1.
To adjust the scale of the open-to-close integrated volatility $IV_{i}$ and the overnight volatility $OV_{i}$, we set  the weight $w$ as 0.75. 
Then the conditional standard deviation of the diffusion process satisfies
\begin{eqnarray*}
&&h_i(\theta) = \omega + \gamma h_{i-1}(\theta) + \alpha\sqrt{IV_{i-1}} + \beta\sqrt{OV_{i-1}},\cr
&&\int_{i-1}^i \sigma_t^2 dt = h^2_i\left(\theta\right)\left(1+d_i\right) = h^2_i\left(\theta\right) + D_i,\cr
&&Y_{i} = h_i(\theta)\epsilon_i,
\end{eqnarray*}
where   $D_i = h^{2}_{i}(\theta)d_i$, $i=1,\ldots,n$, are martingale differences,  $\epsilon_i$'s are i.i.d. random variables, which have the same distribution as that of $Z_{i}\sqrt{1+d_i}$, where $Z_i$'s are i.i.d. standard normal random variables, and the true model parameter $\left(\omega,\gamma,\alpha,\beta\right) = (1,0.1,0.5,0.2)$.
We varied $n$ from 500 to 2000 and $m$ from 100 to 1000.
The whole simulation procedure was repeated 1000 times.

\begin{figure}[!ht]
\centering
\includegraphics[width = 1\textwidth]{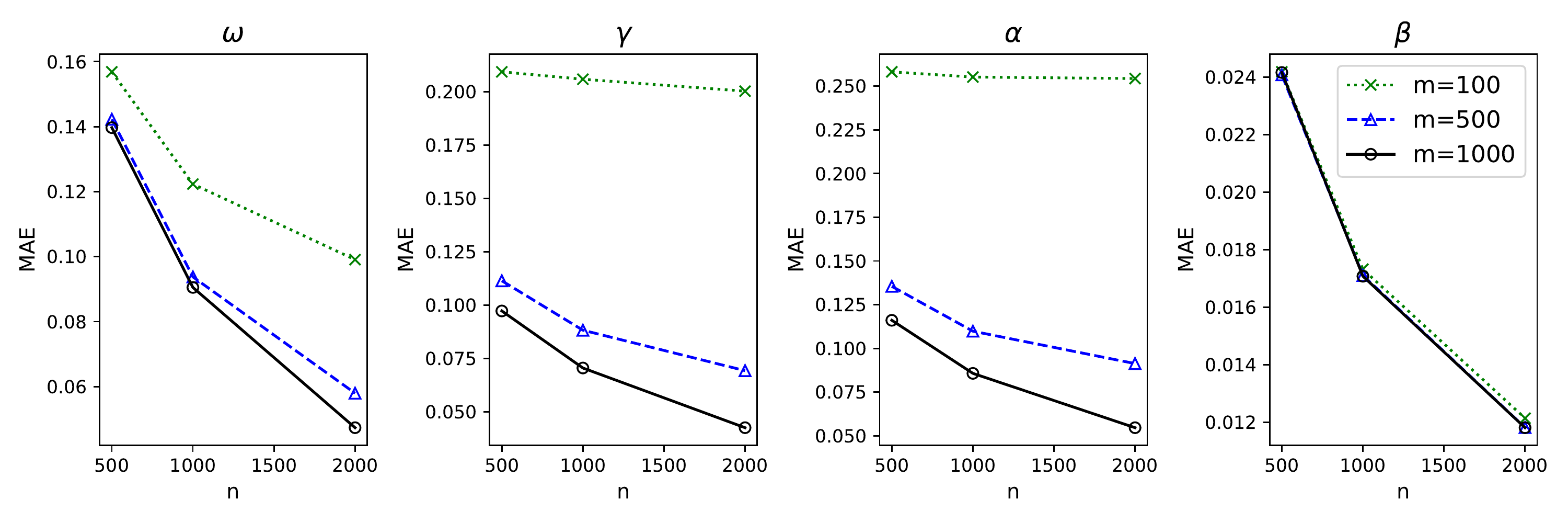}
\caption{MAEs for the first step estimation with $n=500,1000,2000$ and $m=100,500,1000$.} \label{Fig-1}
\end{figure}

\begin{figure}[!ht]
\centering
\includegraphics[width = 0.75\textwidth]{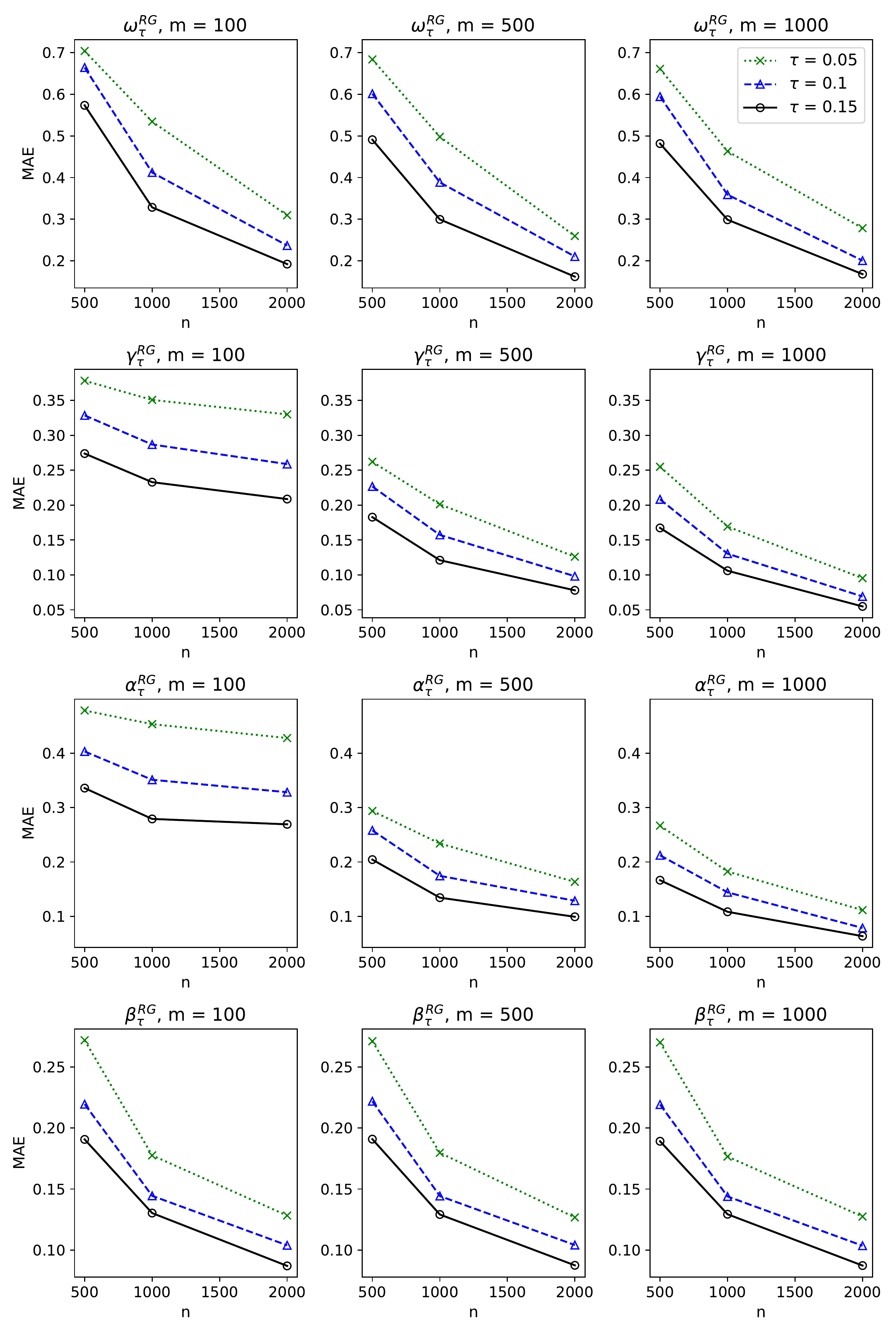}
\caption{MAEs for the proposed  realized GARCH quantile regression two-step estimator $\hat{\theta}_{\tau}^{RG}$ with $n=500,1000,2000$, $m=100,500,1000$, and $\tau=0.05,0.1,0.15$.} \label{Fig-2}
\end{figure}

\begin{figure}[!ht]
\centering
\includegraphics[width = 0.75\textwidth]{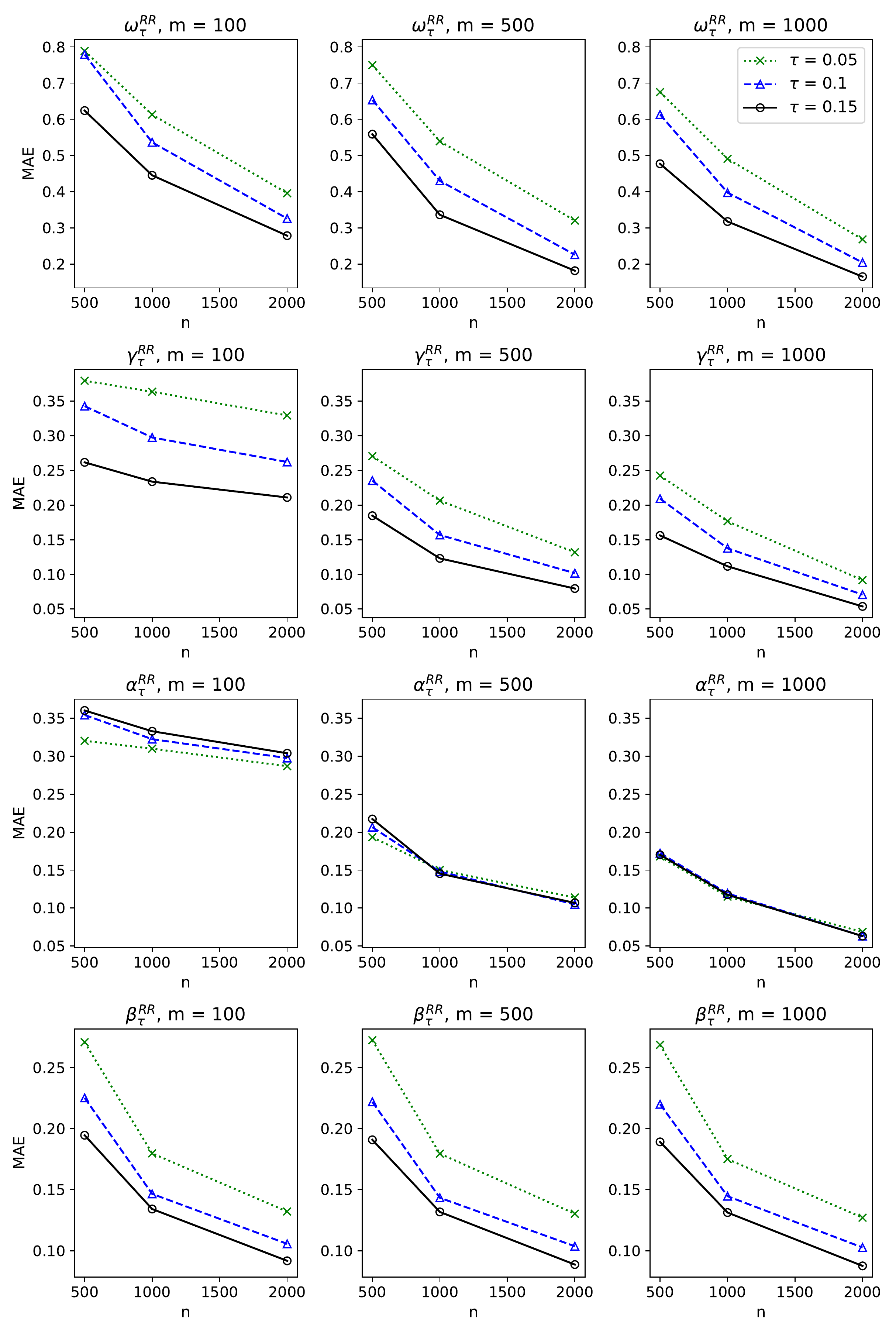}
\caption{MAEs for the proposed  real-realized GARCH quantile regression two-step estimator $\hat{\theta}_{\tau}^{RR}$ with $n=500,1000,2000$, $m=100,500,1000$, and $\tau=0.05,0.1,0.15$.} \label{Fig-3}
\end{figure}

Figure \ref{Fig-1} depicts the estimated mean absolute errors (MAE)  of the first step estimates with $n=500,1000,2000$ and $m=100,500,1000$.
Figures \ref{Fig-2}--\ref{Fig-3} draw the MAEs for the realized GARCH and real-realized GARCH quantile regression  two-step estimates with $n=500,1000,2000$, $m=100,500,1000$ and $\tau=0.05,0.1,0.15$.
Figures \ref{Fig-1}--\ref{Fig-3} show that the MAEs usually decrease as the number of high-frequency observations or low-frequency observations increases.
We note that there is a little effect of increasing the number of high-frequency observations in case of $\hat{\beta}^{RG}, \hat{\beta}^{RR}, \hat{\omega}^{RG}$, and $\hat{\omega}^{RR}$ which are the coefficients of $OV$ and the intercept term in the quantile regression.
This may be because the quanitle regression is based on the low-frequency observations, so high-frequency observations have relatively little effect on the estimation accuracy.
From Figures \ref{Fig-2}--\ref{Fig-3}, we find that the MAEs decrease as $\tau$ increases except $\halpha^{RR}_{\tau}$.
This is because while the other quantile regression parameters decrease in proportion to the decrease of $q_{\tau}$,  $\alpha^{RR}_{\tau}=\alpha q_{\tau}/z_{\tau}$ does not decrease as $\tau$ increases.
These results support the theoretical findings in Section \ref{SEC-3}.

Our main goal in this paper is to predict the conditional quantile.
We therefore investigated the out-of-sample performance of estimating the one-day-ahead conditional quantile.
To predict the one-day-ahead conditional quantile, we employed the proposed two-step estimators, and the conditional quantile can be calculated as follows:
\begin{eqnarray*}
&& \hat{Q}_{\tau,n+1}^{RG} =  \homega_\tau^{RG} + \hat{\gamma}_\tau^{RG} \hat{h}_{n}(\htheta) + \halpha_\tau^{RG} \sqrt{RV_n} + \hbeta_\tau^{RG} \sqrt{OV_n} ,\\
&& \hat{Q}_{\tau,n+1}^{RR} =  \homega_\tau^{RR} + \hat{\gamma}_\tau^{RR} \hat{h}_{n}(\htheta) + \halpha_\tau^{RR} \hat{RQ}_{\tau,n} + \hbeta_\tau^{RR} \sqrt{OV_n}.
\end{eqnarray*}
The true conditional quantile is $h_{n+1}(\theta)q_{\tau}$, where $q_{\tau}$ is the $\tau$-quantile value of $\epsilon_n$. We calculated the value of $q_{\tau}$ by Monte Carlo method.
For comparisons, we consider  the QGARCH \citep{xiao2009conditional} and the realized CAViaR \citep{vzikevs2015semi}, which are the two-step estimation using low-frequency observations for the quantile regression based on the GARCH model and the conditional autoregressive quantile regression using high-frequency observations, respectively.
Specifically, we chose the GARCH(1,1) with the absolute value of daily returns as the innovation for the QGARCH, and the square root of realized volatilities and the absolute value of daily returns were utilized for the realized CAViaR.

\begin{figure}[!ht]
\centering
\includegraphics[width = 0.8\textwidth]{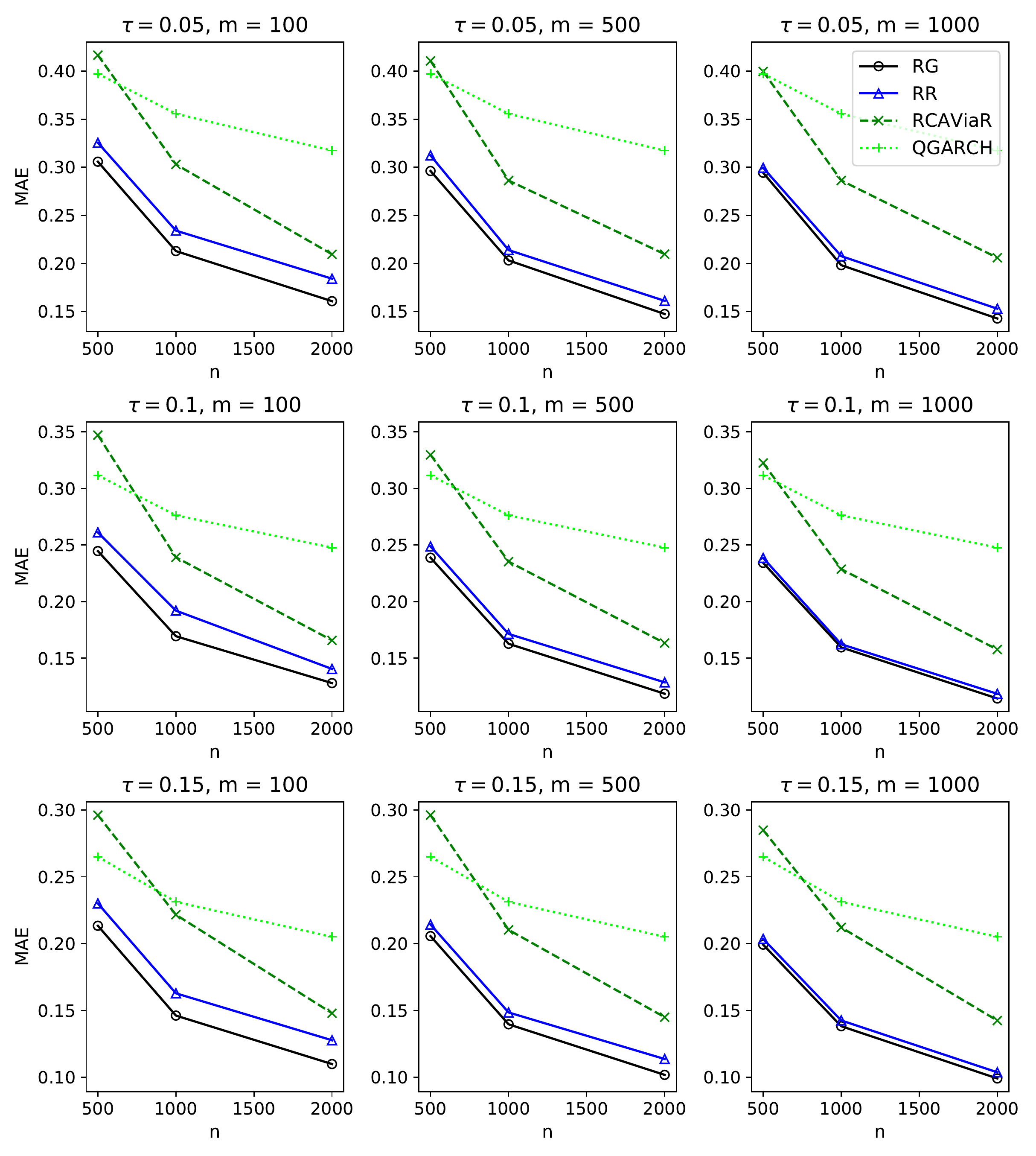}
\caption{MAEs for the conditional quantile with RG, RR, RCAViaR and QGARCH two-step estimators with $n=500,1000,2000$, $m=100,500,1000$, and $\tau=0.05,0.1,0.15$.}
 \label{Fig-4}
\end{figure}

Figure \ref{Fig-4} depicts the estimated  MAEs of the one-day-ahead conditional quantile with the realized GARCH quantile regression (RG), real-realized GARCH quantile regression (RR), realized CAViaR (RCAViaR), and QGARCH two-step estimators with  $n=500,1000,2000$, $m=100,500,1000$, and $\tau=0.05,0.1,0.15$.
From Figure \ref{Fig-4}, we observe that the MAEs in the RG, RR and RCAViaR cases decrease as the number of low-frequency observations or high-frequency observations increases.
The  RG and RR models exhibit better performance than RCAViaR and QGARCH, and when $n$ is large, the QGARCH shows the worst performance. 
One of the possible explanations is that the QGARCH does not include the high-frequency information, so it cannot explain the volatility dynamics well. 
On the other hand, since the RCAViaR includes the high-frequency information,  the RCAViaR is able to capture the volatility dynamics.
However, its estimation procedure is relatively complicated compared to the proposed two step estimation method, which may cause some estimation errors. 
When comparing the RG and RR models, the RG model shows slightly better performance.

\section{An empirical study} \label{SEC-5}

We applied the proposed realized GARCH quantile regression model and the real-realized GARCH quantile regression model to measuring the conditional quantile of the real high-frequency trading data.
We selected the top 20 large trading volume stocks among the S\&P 500 compositions.
To minimize the effect of the micro-structure noise, we used the 5 minutes intraday trading data for the selected stocks from January 2010 to December 2016, 1758 trading days in total.
We obtained the data from Wharton Data Service (WRDS) system.
We defined the open-to-close period from 9:30 to 16:00, the close-to-open period from 16:00 to the following-day 9:30, and the one-day unit period as the close-to-close period.
We utilized the log-prices for estimating the conditional quantile of the daily log-returns.

To predict one-day-ahead conditional quantile, we used  the RG, RR, RCAViaR, and QGARCH models defined in Section \ref{SEC-4}, and non-parametric sample quantile (SQ).
We set the in-sample period as 500 days, and using the rolling window scheme, we  predicted the one-day-ahead conditional quantile for the last 1258 days with $\tau = 0.01, 0.03, 0.05, 0.1, 0.15$.
For relative comparisons of the models, we used the quantile loss function \citep{koenker1978regression} as follows
\begin{equation*}
	L(\bY,\hat{\bQ}_\tau) = \frac{1}{n} \sum_{i=1}^{n} \rho _{\tau} (e_{\tau,i}) = \frac{1}{n} \sum_{i=1}^{n} e_{\tau,i} ( \tau - \I (e_{\tau,i} <0)),
\end{equation*}
where $\bY = (Y_1, \ldots, Y_n)^\top$ are close-to-close log-returns, $\hat{\bQ}_\tau = (\hat{Q}_{\tau,1}, \ldots, \hat{Q}_{\tau,n})^\top$ are estimated conditional quantiles, and $e_{\tau,i}=Y_i - \hat{Q}_{\tau,i}$.
For each individual stock and quantile level $\tau$, we calculated the quanitle loss for each model, and divided the loss by the loss of the RG model to check relative performance.
We call this the relative loss.
We note that  the relative loss of the RG model is 1.

\begin{figure}[!ht]
\centering
\includegraphics[width = 1.0\textwidth]{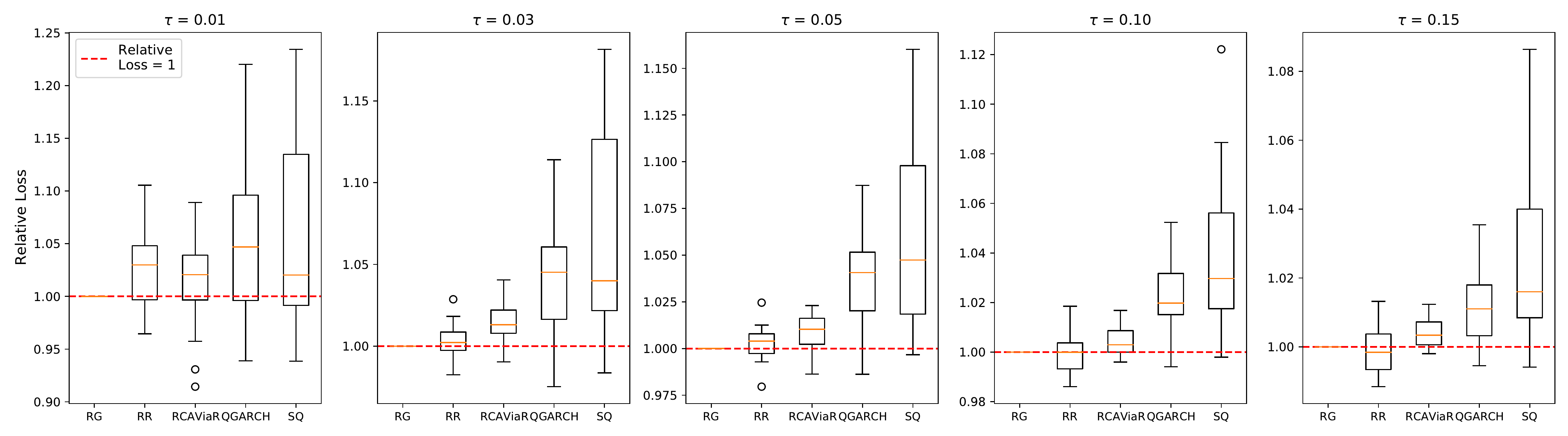}
\caption{Boxplots for the relative losses with the RG, RR, RCAViaR and QGARCH models and SQ for the 20 individual stocks against varying $\tau = 0.01,0.03,0.05,0.1,0.15$.} \label{Fig-6}
\end{figure}

\begin{table}[!ht]
\caption{Average rank of relative losses for the RG, RR, RCAViaR and QGARCH models and SQ over the 20 individual stocks against varying $\tau = 0.01,0.03,0.05,0.1,0.15$. In the parenthesis, we report the number of the first rank of relative losses among competitors.
}\label{Table-2}
\centering
\scalebox{1}{
\begin{tabular}{llllll}
\hline
$\tau$ & \multicolumn{1}{l}{RG} & \multicolumn{1}{l}{RR} & \multicolumn{1}{l}{RCAViaR} & \multicolumn{1}{l}{QGARCH} &\multicolumn{1}{l}{SQ} \\ \hline
0.01   & 2.65 (8)               & 2.80 (1)               & 2.70 (5)                   & 3.15 (2)                   & 3.70 (4)               \\
0.03   & 1.90 (9)               & 1.80 (6)               & 2.95 (3)                   & 3.90 (2)                   & 4.45 (0)               \\
0.05   & 1.60 (12)              & 2.40 (5)               & 2.85 (2)                   & 3.80 (1)                   & 4.35 (0)               \\
0.1   & 2.20 (5)               & 1.65 (9)               & 2.50 (4)                   & 4.20 (1)                   & 4.45 (1)               \\
0.15   & 1.80 (7)               & 2.25 (11)              & 3.00 (1)                   & 3.50 (1)                   & 4.45 (0)               \\ \hline
\end{tabular}
}
\end{table}

Figure \ref{Fig-6} draws boxplots for the relative losses with the RG, RR, RCAViaR and QGARCH models and SQ over the 20 individual stocks against varying $\tau = 0.01, 0.03,0.05, 0.1, 0.15$.
The horizontal red dot lines in Figure \ref{Fig-6} indicate the relative loss 1, so the model having the box over the red dot line performs worse than the RG model.
Table \ref{Table-2} reports the average rank and the number of first rank of the relative loss for the RG, RR, RCAViaR and QGARCH models and sample quantile over the 20 individual stocks.
From Figure \ref{Fig-6} and Table \ref{Table-2}, we find that the parametric models show perform better than the non-parametric sample quantile.
Moreover, the RG, RR and RCAViaR models that utilize high-frequency information show better performance than QGARCH model which uses only low-frequency information.
When comparing the models using high-frequency information, the proposed RG and RR model show the best performance for $\tau = 0.01,0.03,0.05$ and $\tau = 0.1,0.15$, respectively.
From this result, we can conjecture that the proposed model can account for the market quantile dynamics via incorporating the high-frequency information and the simple two step estimation procedure reduces the estimation errors.

To backtest the estimated conditional quantile, we conducted  hypothesis tests as follows.
 We first calculated $I_i = \mathbbm{1}(Y_i - Y_{i-1} < \hat{Q}_{\tau,i})$, where $\mathbbm{1}(\cdot)$ is an indicator function and $\hat{Q}_{\tau,i}$ is predicted conditional quantile with quantile level $\tau$. 
 Then we conducted hypothesis tests based on the assumptions that  $I_i - \tau$ has mean 0 and is a martingale difference sequence. 
 For example, the following three test statistics are calculated to carry out the hypothesis tests. 
 The first one is the likelihood ratio unconditional coverage (LRuc) test proposed by \citet{kupiec1995techniques}:
\begin{equation*}
    LR_{uc} = -2\log\left(\frac{\tau^x(1-\tau)^{n-x}}{\left(1-x/n\right)^{n-x}\left(x/n\right)^x}\right),
\end{equation*}
where $n$ is the number of predicted conditional quantiles and $x = \sum_{i=1}^{n}I_i$. 
The LRuc is based on the independent assumption of $I_i$'s, thus, it cannot explain the dynamic structure.
The second one is the likelihood ratio conditional coverage (LRcc) test proposed by \citet{christoffersen1998evaluating}:
\begin{equation*}
    LR_{cc} = -2\log\left(\frac{\tau^{x}(1-\tau)^{n-x}}{L(\hat{\Pi} ; I_1,\ldots,I_n)}\right),
\end{equation*}
where $L(\Pi ; I_1,\ldots,I_n) = \pi_{01}^{n_{01}}(1-\pi_{01})^{n_{00}}\pi_{11}^{n_{11}}(1-\pi_{11})^{n_{10}}$,$\pi_{ij} = P(I_{d+1}=j|I_{d}=i)$, $n_{ij}$ is the number of $j$ outcomes after $i$ outcome, and $\hat{\Pi}$ is the maximum likelihood estimator.
The LRcc test considers the one lagged relationship. 
The third one is the dynamic quantile (DQ) test, proposed by \citet{engle2004caviar}, with the first $L$ lagged $I_i$'s and the VaR forecast.
The DQ test considers some dynamic structure for $L$ lagged variables.
In this paper, we chose $L=4$. 
 Details of the test statistic can be found in \citet{engle2004caviar}. 
We conducted hypothesis tests with 20 individual stocks.

\begin{figure}[!ht]
\centering
\includegraphics[width = 1.0\textwidth]{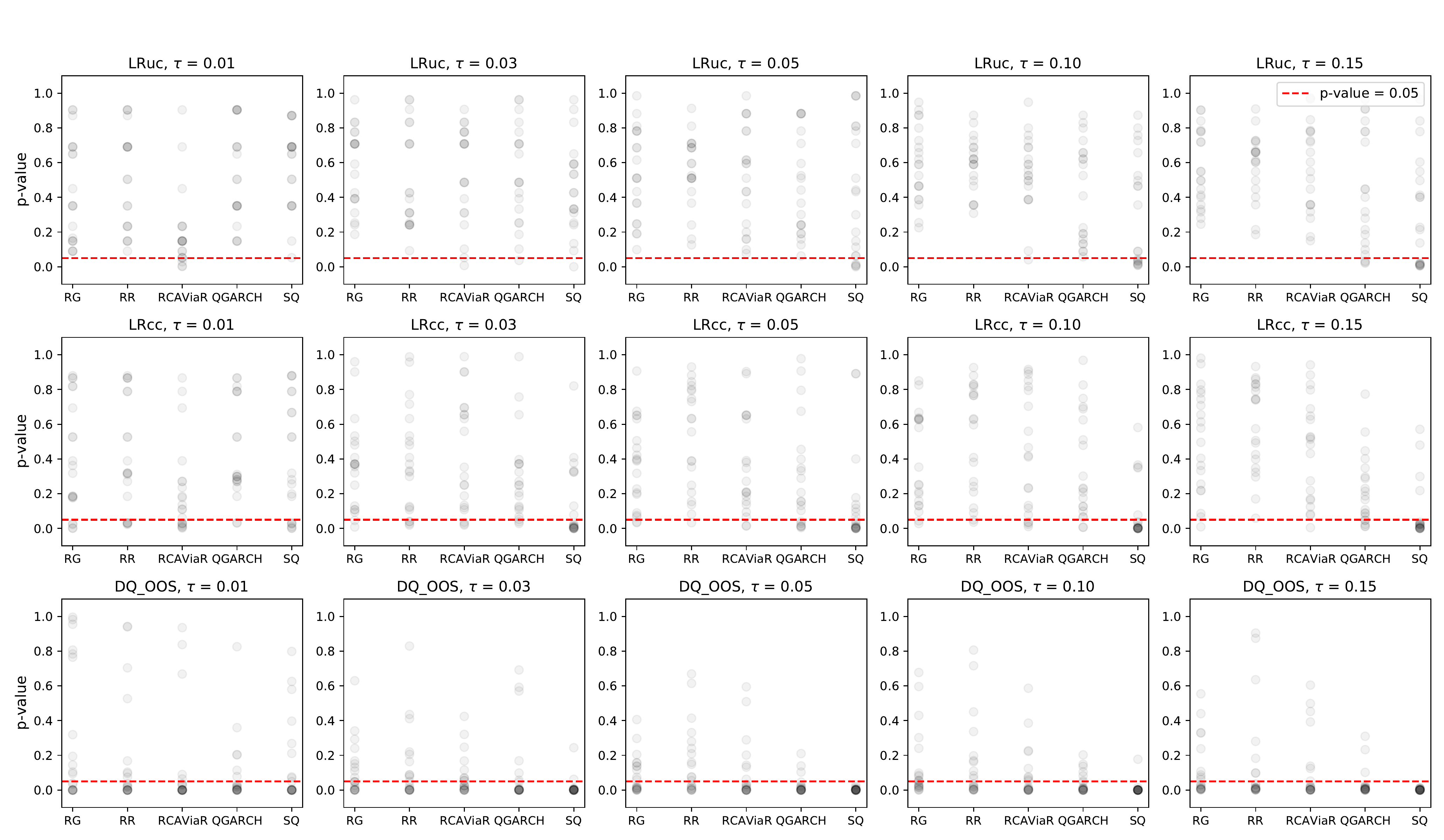}
\caption{P-values of the LRuc, LRcc, and DQ tests with the RG, RR, RCAViaR and QGARCH models and SQ for 20 individual stocks against varying $\tau = 0.01,0.03,0.05,0.1,0.15$.} \label{Fig-5}
\end{figure}

\begin{table}[!ht]
\caption{The number of individual stocks whose p-values are below 0.05.
 }\label{Table-1}
\centering
\scalebox{1}{
\begin{tabular}{clrrrrr}
\hline
\multicolumn{1}{l}{} & $\tau$ & \multicolumn{1}{l}{RG} & \multicolumn{1}{l}{RR} & \multicolumn{1}{l}{RCAViaR} & \multicolumn{1}{l}{QGARCH} & \multicolumn{1}{l}{SQ} \\ \hline
					 & 0.01   & 0                      & 0                      & 3                          & 0                          & 0                      \\
					 & 0.03   & 0                      & 0                      & 1                          & 1                          & 1                      \\
LRuc                 & 0.05   & 0                      & 0                      & 0                          & 0                          & 4                      \\
					 & 0.1    & 0                      & 0                      & 1                          & 0                          & 8                      \\
					 & 0.15   & 0                      & 0                      & 0                          & 3                          & 8                      \\ \hline
					 & 0.01   & 4                      & 5                      & 7                          & 2                          & 6                      \\
					 & 0.03   & 2                      & 4                      & 3                          & 2                          & 12                     \\
LRcc                 & 0.05   & 2                      & 1                      & 2                          & 6                          & 12                     \\
					 & 0.1    & 2                      & 2                      & 4                          & 2                          & 15                     \\
					 & 0.15   & 1                      & 0                      & 1                          & 5                          & 16                     \\ \hline
					 & 0.01   & 9                      & 12                     & 15                         & 14                          & 12                     \\
					 & 0.03   & 11                     & 11                     & 11                         & 14                         & 18                     \\
DQ                   & 0.05   & 10                     & 8                      & 13                         & 17                         & 20                     \\
					 & 0.1    & 10                     & 10                     & 11                         & 14                         & 19                     \\
					 & 0.15   & 11                     & 13                     & 13                         & 17                         & 20                     \\ \hline
\end{tabular}
}
\end{table}	

Figure \ref{Fig-5} draws p-value scatter plots of the LRuc, LRcc, and DQ tests with the RG, RR, RCAViaR, and QGARCH models and sample quantiles for 20 individual stocks against varying $\tau = 0.01,0.03,0.05,0.1,0.15$. 
When several points are overlapped, they are displayed in dark.
The horizontal red dot lines in Figure \ref{Fig-5} indicate the p-value 0.05.
Therefore, the model having many points under the red dot line fails to predict conditional quantile.
Table \ref{Table-1} reports the number of individual stocks whose p-values are below 0.05.
From Figure \ref{Fig-5} and Table \ref{Table-1}, we find that the parametric models show better performance than the non-parametric sample quantile.
It may be because the quantile has some time series dynamic structure.
When comparing the parametric models, the proposed RG and RR models show slightly better performance than others. 
For the DQ test, the RG, RR and RCAViaR models that utilize high-frequency information show better performance than QGARCH model that uses only low-frequency information.
From these results, we may conjecture that incorporating the realized quantities such as realized volatility and realized quantile helps to capture the quanitle dynamics.

 \section{Conclusion}
 In this paper, we propose quantile regression models with the realized quantities such as realized volatility and realized quantile, based on the realized GARCH models. 
 For example, the realized GARCH quantile regression is based the overnight GARCH-It\^o model and incorporates the past conditional GARCH volatility, open-to-close realized volatility, and overnight return as the explanatory variables. 
In contrast, under the self-similarity condition, we propose the real-realized GARCH quantile regression model, which employs the realized quantile estimator as the explanatory variables. 
In order to reduce the complexity of estimation procedure, we introduce the two-step estimation procedure and show its asymptotic properties. 
From the empirical study, we find that incorporating the realized quantities such as realized volatility and realized quantile helps to capture the quanitle dynamics.

To incorporate the realized quantile \citep{dimitriadis2021realized}, we assume the self-similarity condition, which is often violated in the real data analysis. 
Thus, it is interesting to develop an estimation procedure of the realized quantile, which is robust to the self-similarity condition.
Furthermore, in the financial data analysis, we often observe the leverage effect, and several empirical studies with the realized volatility showed that consider the leverage effect helps account for market dynamics \citep{chun2021state, hansen2016exponential}. 
In this point of view, incorporating the leverage effect in the quantile regression modeling may also help to explain the quantile dynamics.
 We leave these interesting topics for future study.

 \section{Proofs} \label{SEC-Proof}
 
  We first fix some notation. 
 For any given vector $b= (b_i) _{i=1,\ldots, k}$, we define $\|b \|_{\max}=\max_i |b_i|$ and $\| b\|_a = (\sum_{i=1}^k |b_i| ^a )^{1/a}$ for some constant $a>0$.  
 Let $C$'s be positive generic constants whose values are
independent of $\theta$, $n$, and $m$  and may change from occurrence to occurrence.

 \subsection{Proof of Theorem \ref{Thm-1}}
\textbf{Proof of Theorem \ref{Thm-1}.}   
First, we consider \eqref{Thm1-result1}. 
By the mean value theorem and Taylor expansion, there exists $\tilde{\theta}$ between $\hat{\theta}$ and $\theta_0$ such that
\begin{equation*}
	 \frac{\partial \hat{L}_{n,m} (\hat{\theta})  }{\partial \theta }  -    \frac{\partial \hat{L}_{n,m} (\theta_0)  }{\partial \theta }   =  \frac{\partial^2 \hat{L}_{n,m} (\tilde{\theta})  }{\partial \theta \partial \theta ^{\top} }   (\hat{\theta} - \theta_0)  .
\end{equation*}
Similar to the proofs of Theorem 1 \citep{song2021volatility}, we can show 
\begin{equation*}
 \frac{\partial \hat{L}_{n,m} (\theta_0)  }{\partial \theta } = \frac{1}{n}  \sum_{i=1}^n \frac{D_i ^L}{h_i^4 (\theta_0) } \frac{\partial h_i^2 (\theta_0)}{ \partial \theta}  +O_p (m^{-1/2}) 
\end{equation*}
and
\begin{equation*}
 - \frac{\partial^2 \hat{L}_{n,m} (\tilde{\theta})  }{\partial \theta \partial \theta ^{\top} }  \overset{p}{\to} B_1.
\end{equation*}
Thus, we have
\begin{equation*}
	\sqrt{n} ( \hat{\theta} - \theta_0 )=  \frac{B_1 ^{-1}}{\sqrt{n}}  \sum_{i=1}^n \frac{D_i ^L }{h_i^4 (\theta_0) } \frac{\partial h_i^2 (\theta_0)}{ \partial \theta} +O_p(n^{1/2} m^{-1/2}) + o_p(1).
\end{equation*}

The result of \eqref{Thm1-result1},  the martingale central limit theorem and Cram\'er-Wold device imply the statement  \eqref{Thm1-result2} immediately. 
\endpf

    \subsection{Proof of Theorem \ref{Thm-2}}
    \textbf{Proof of Theorem \ref{Thm-2}.}
    To easy the notation, we denote $\hat{\theta}_{\tau} ^{RG}$ by $\hat{\theta}_\tau$.
We define
\begin{eqnarray*}
&&\hat{h}_{n} ( \theta ,a_0)= \omega   + \gamma  \hat{h}_{i-1}( \theta, a_0) + \alpha   \sqrt{RV_{n-1}} +  \beta  \sqrt{ OV _{n-1}}, \cr
&&h_{n} ( \theta ,a_0)=  \omega   +  \gamma h_{i-1}(  \theta, a_0) + \alpha   \sqrt{IV_{n-1}} +  \beta  \sqrt{ OV _{n-1}},  
\end{eqnarray*}    
where $\hat{h}_1 (\theta) = a_0$ and $h_1 (\theta) = a_0$.   
Denote  the true initial value $h_1 (\theta_0)$ by $h_0$.
Let
\begin{eqnarray*}
	&&\hat{G}_{n,m} (\theta_\tau,\theta, a_0 ) = \frac{1}{n- 1}  \sum_{i=2}^n \left  \{ \tau - \I (Y_i   < \theta_\tau^{\top} \hat{A}_{m,i}  (\theta, a_0) ) \right \} \hat{A}_{m,i} (\theta, a_0),  \cr
   &&G _n (\theta_\tau, \theta ) = \frac{1}{n- 1}  \sum_{i=2}^n \left  \{ \tau - \I (Y_i   < \theta_\tau^{\top} A_i (\theta)  ) \right \} A_i (\theta) , \cr
   && G  (\theta_\tau, \theta) =    E \[   \left  \{ \tau - F_{Y|X}  ( \theta_\tau^{\top} A_i (\theta)  ) \right \} A_i (\theta)   \],
\end{eqnarray*}
where $\hat{A}_{m,i} (\theta, a_0) = ( 1, \hat{h} _{i-1} ( \theta ,a_0),  \sqrt{ RV_{i-1}} , \sqrt{OV _{i-1}}) ^{\top}$,  $A_i (\theta) = ( 1, h _{i-1} (\theta, h_0),  \sqrt{ IV_{i-1}} , \sqrt{OV_{i-1}}) ^{\top}$.
Then the quantile regression estimator $\hat{\theta}_{\tau}$ is the same as 
$$
\hat{\theta}_{\tau}  = \arg \min _{\theta_{\tau}}  \|\hat{G}_{n,m} (\theta_\tau, \hat{\theta}, a_0  )\|_2, 
$$
and the true parameter is 
$$
\theta_{\tau 0}  =  \arg \min _{\theta_{\tau}}  \| G   (\theta_\tau, \theta_0 )\|_2.
$$
We note that $ G   (\theta_{\tau 0}, \theta_0 ) =0$.

First, we show the consistency of $\hat{\theta}_{\tau}$. 
Since $\hat{\theta}$ is the consistent estimator of $\theta_0$, it is enough to show the statement under $\| \theta - \theta_0\|_1 \leq \delta_n$, where $\delta_n = o(1)$. 
We have
\begin{eqnarray*}
 && \|\hat{G}_{n,m} (\theta_\tau,\theta, a_0 ) - G _n (\theta_\tau, \theta )  \| _1  \cr
 &&\leq   \frac{1}{n- 1}   \sum_{i=2}^n  \|  \hat{A}_{m,i} (\theta, a_0) - A_i (\theta) \|_1 \cr
 && \qquad   +    \frac{1}{n- 1}  \left \|    \sum_{i=2}^n    \left  \{ \I (Y_i   < \theta_\tau^{\top} A_i (\theta)  ) - \I (Y_i   < \theta_\tau^{\top} \hat{A}_{m,i}  (\theta, a_0) ) \right \} A_i (\theta)  \right \|_1  .
\end{eqnarray*}
We have
\begin{eqnarray}  \label{thm1-eq-01}
	  &&  \sup_{\theta}\|  \hat{A}_{m,i} (\theta, a_0) - A_i (\theta) \|_1 \cr
	   &&\leq   \sup_{\theta}  |\hat{h} _{i-1} ( \theta ,a_0) - h _{i-1} ( \theta)| + | \sqrt{ RV_{i-1}}- \sqrt{ IV_{i-1}} | \cr
	  	&&\leq C  \sum_{j=1}^{i-2} \gamma_u ^{j-1}   | \sqrt{ RV_{i-1-j}} - \sqrt{ IV_{i-1-j}} |  +   \gamma_u ^{i-2} |a_0 -h_0| + | \sqrt{ RV_{i-1}}- \sqrt{ IV_{i-1}} | \cr
	  	&&= O_p(m^{-1/2} ) +   \gamma_u ^{i-2} |a_0 -h_0|.
\end{eqnarray}
Thus, we have
\begin{equation}  \label{eq00001}
	 \sup_{\theta} \frac{1}{n- 1}   \sum_{i=2}^n  \|  \hat{A}_{m,i} (\theta, a_0) - A_i (\theta) \|_1 = O_p (m^{-1/2} + n^{-1}).
\end{equation}
We have
\begin{eqnarray*}
	 \I (Y_i   < \theta_{\tau } ^{\top} A_i (\theta)  ) &=&    \I (Y_i   < \theta_{\tau } ^{\top} A_i (\theta_0 )   +  \theta_{\tau } ^{\top} (A_i (\theta) - A_i (\theta_0 ) )  )  \cr
	 &=& \I (\epsilon _i   <  \theta_{\tau } ^{\top} A_i (\theta_0 )/ h_i (\theta_0)    +  \theta_{\tau  } ^{\top} (A_i (\theta) - A_i (\theta_0 ) )  / h_i (\theta_0)    ) 
\end{eqnarray*}
and 
\begin{eqnarray*}
	   \I (Y_i   < \theta_{\tau } ^{\top} \hat{A}_{m,i}  (\theta, a_0) )  &=&  \I (Y_i   < \theta_{\tau } ^{\top} A_i (\theta_0 )  + \theta_{\tau  } ^{\top}  (\hat{A}_{m,i} (\theta, a_0)  - A_i (\theta_0) ) ) )  \cr
	 &=& \I (\epsilon_i   < \theta_{\tau } ^{\top} A_i (\theta_0 )/h_i (\theta_0)     + \theta_{\tau  } ^{\top}  (\hat{A}_{m,i} (\theta, a_0)  - A_i (\theta_0) ) ) / h_i (\theta_0)  ).
\end{eqnarray*}
Thus, we have
\begin{eqnarray} \label{eq-000000}
&&\sup_{\| \theta  -\theta_0 \|_1 \leq \delta_n }  |   \I (Y_i   < \theta_{\tau  } ^{\top} A_i (\theta)  )  -  \I (Y_i   < \theta_{\tau } ^{\top} \hat{A}_{m,i}  (\theta, a_0) ) | \cr
  && \leq \sup_{\| \theta  -\theta_0 \|_1 \leq \delta_n } \I \( \left | \epsilon_i -  \frac{  \theta_{\tau } ^{\top} A_i (\theta_0 )}{  h_i (\theta_0)} \right | < \frac{  | \theta_{\tau  } ^{\top} (A_i (\theta) - A_i (\theta_0 ))  |   \vee | \theta_{\tau  } ^{\top} ( \hat{A}_{m,i} (\theta, a_0)  - A_i (\theta_0) )  |   }{ h_i (\theta_0) }   \) .
\end{eqnarray}
By \eqref{thm1-eq-01}, for $i \geq C \log n$, we have  $\sup_{\| \theta  -\theta_0 \|_1 \leq \delta_n } \max(| \theta_{\tau  } ^{\top} (A_i (\theta) - A_i (\theta_0 ))  |   , | \theta_{\tau  } ^{\top} ( \hat{A}_{m,i} (\theta, a_0)  - A_i (\theta_0) )  | ) / \theta_{ 0} ^{\top} A_i (\theta_0)  = o_p(1)$, so we have
\begin{eqnarray*} 
&&E \[ \sup_{\| \theta  -\theta_0 \|_1 \leq \delta_n } |    \I (Y_i   < \theta_{\tau  } ^{\top} A_i (\theta)  )  -  \I (Y_i   < \theta_{\tau  } ^{\top} \hat{A}_{m,i}  (\theta, a_0) ) |  \] \cr
&&\leq E \[  \sup_{\| \theta  -\theta_0 \|_1 \leq \delta_n } \I \( \left | \epsilon_i -  \frac{  \theta_{\tau } ^{\top} A_i (\theta_0 )}{  h_i (\theta_0)} \right | < \frac{  | \theta_{\tau  } ^{\top} (A_i (\theta) - A_i (\theta_0 ))  |   \vee | \theta_{\tau  } ^{\top} ( \hat{A}_{m,i} (\theta, a_0)  - A_i (\theta_0) )  |   }{ h_i (\theta_0) }   \)  \] \cr
&&\leq \sup_{\| \theta  -\theta_0 \|_1 \leq \delta_n } E \[   \I \( \left | \epsilon_i -  \frac{  \theta_{\tau } ^{\top} A_i (\theta_0 )}{  h_i (\theta_0)} \right | < \frac{  | \theta_{\tau  } ^{\top} (A_i (\theta) - A_i (\theta_0 ))  |   \vee | \theta_{\tau  } ^{\top} ( \hat{A}_{m,i} (\theta, a_0)  - A_i (\theta_0) )  |   }{ h_i (\theta_0) }   \)  \] \cr
&&\leq o(1),
\end{eqnarray*}
where the second inequality is by the fact that $I ( \left | \epsilon_i -  \frac{  \theta_{\tau } ^{\top} A_i (\theta_0 )}{  h_i (\theta_0)} \right |  \leq \cdot )$ is a momotone function.  
Thus,  we have
\begin{eqnarray*}
	 	&&\sup_{\| \theta  -\theta_0 \|_1 \leq \delta_n }   \frac{1}{n- 1}  \left \|    \sum_{i=2}^n    \left  \{ \I (Y_i   < \theta_{\tau } ^{\top} A_i (\theta)  ) - \I (Y_i   < \theta_{\tau } ^{\top} \hat{A}_{m,i}  (\theta, a_0) ) \right \} A_i (\theta)  \right \|_1  \cr
	 	&&  = o_p(1) +  O_p\( \frac{  \log n }{n} \) = o_p(1) 
\end{eqnarray*}
and together with \eqref{eq00001},  
\begin{equation} \label{Thm2-eq-r01}
	 \sup_{\| \theta  -\theta_0 \|_1 \leq \delta_n } \| \hat{G}_{n,m} (\theta_{\tau } ,\theta, a_0 ) - G _n (\theta_{\tau  } , \theta )  \| _1 = o_p(1) .
\end{equation}

Consider $\|G _n (\theta_\tau, \theta  ) - G(\theta_\tau, \theta ) \|_2$. 
For any given $\theta_{\tau}$ and $\theta$,   we can show
\begin{equation*}
	\|G _n (\theta_\tau, \theta  ) - G(\theta_\tau, \theta ) \|_2 = O_p(n^{-1/2}).
\end{equation*}
Now, it is enough that $G _n (\theta_\tau, \theta  ) $ is stochastically equicontinuous under $\| \hat{\theta} -\theta_0\|_1 \leq \delta_n$. 
Let
$$
m_j (A_i ,\theta_\tau, \theta) =  \left  \{   \I (Y_i   < \theta_{\tau }^{\top} A_i (\theta ))  - \tau   \right \} A_{ji} (\theta ) ,
$$
where $A_{ji} (\theta ) $ is the $j$th element of $A_{i} (\theta )$. 
We have
\begin{eqnarray}\label{Thm2-eq001}
	&& E \[ \sup_{ \theta^{\prime}, \theta:   \| \theta - \theta_0 \|_1  \leq \delta_n }    | m_j (A_i ,\theta_{\tau  }, \theta ^{\prime} )- m_j (A_i ,\theta_{\tau  } , \theta )|  \] \cr
	&&\leq   E \[ \sup_{ \theta^{\prime}, \theta:   \| \theta - \theta_0 \|_1  \leq \delta_n }   |  A_{ji} (\theta ^{\prime})  - A_{ji} (\theta  ) |  \] \cr
	&& \quad  +   E \[ \sup_{ \theta^{\prime}, \theta:   \| \theta - \theta_0 \|_1  \leq \delta_n }   \left |    \I (Y_i   < \theta_{\tau}^{\top} A_i (\theta ^{\prime} ))  -  \I (Y_i   < \theta_{\tau }^{ \top} A_i  (\theta  ) )    \right |   | A_{ji} (\theta  ) |   \]  \cr
	&& \leq C \delta_n + C  E \[ \sup_{ \theta^{\prime}, \theta:   \| \theta - \theta_0 \|_1  \leq \delta_n }   \left |    \I (Y_i   < \theta_{\tau   }^{\top} A_i (\theta ^{\prime}  ))  -  \I (Y_i   < \theta_{\tau  }^{\top} A_i  (\theta   ) )    \right |  \] \cr
	&& \leq C    E \[ \left |    F_{Y|X}  (  \theta_{\tau  }^{\top} A_i (\theta _ 0 + \delta_n ))  -   F_{Y|X}(  \theta_{\tau  }^{\top} A_i  (\theta _ 0 - \delta_n ) )    \right |  \]  + C \delta_n \cr
	&& \leq C    \delta_n, 
\end{eqnarray}
where the third inequality is by the fact that $I(Y_t < \cdot)$ is a monotone function, and the last inequality is due to Assumption \ref{assumption-2}(a). 
Thus, $G _n (\theta_\tau, \theta  ) $ is stochastically equicontinuous.
Therefore, by Theorem 1 \citep{andrews1992generic}, we have
\begin{equation*}
		 \sup_{\| \theta  -\theta_0 \|_1 \leq \delta_n } \|G _n (\theta_{\tau } , \theta  ) - G(\theta_{\tau }, \theta ) \|_2 = o_p(1) ,
\end{equation*}
and, by \eqref{Thm2-eq-r01}, we have, for any given $\theta_\tau$, 
 \begin{equation} \label{Thm2-eq-r02}
	  \sup_{\| \theta  -\theta_0 \|_1 \leq \delta_n } \|\hat{G} _{n,m} (\theta_{\tau }, \theta,a_0  ) - G(\theta_{\tau }, \theta ) \|_2 = o_p(1). 
\end{equation}

Now, we show 
$$
  \sup_{\theta_{\tau} \in \Theta_{\tau}, \| \theta  -\theta_0 \|_1 \leq \delta_n }  \|\hat{G} _{n,m} (\theta_{\tau }, \theta,a_0  ) - G(\theta_{\tau }, \theta ) \|_2 = o_p(1), 
$$
where $\Theta_{\tau}$ is the sample space of $\theta_{\tau}$. 
Since $\Theta_{\tau}$ is a compact set,  without loss of generality, we show the statement under $\| \theta_{\tau} \|_1 \leq M$ for some positive $M$. 
Decompose $\{ \|\Delta \|_1  \leq M \}$ into cubes based on the grid $(j_1 d M, \ldots, j_k d M)$, where $j_i =0 ,\pm1, \ldots, \pm [1/d]+1$, and $d$ is a fixed positive number, and denote the lower vertex of the cube that contain $\theta_{\tau}$ by $L(\theta_\tau)$.
Let $T (\theta_{\tau }, \theta) = \hat{G} _{n,m} (\theta_{\tau }, \theta,a_0  ) - G(\theta_{\tau }, \theta )$. 
Then we have
\begin{eqnarray*}
  && \sup_{\| \theta_{\tau}\|_1 \leq M , \| \theta  -\theta_0 \|_1 \leq \delta_n } \|T (\theta_{\tau }, \theta) \|_2  \cr
  &&\leq \sup_{\| \theta_{\tau}\|_1 \leq M ,  \| \theta  -\theta_0 \|_1 \leq \delta_n } \| T ( L (\theta_{\tau }) , \theta) \|_2  + \sup_{\| \theta_{\tau}\|_1 \leq M , \| \theta  -\theta_0 \|_1 \leq \delta_n } \| T ( L (\theta_{\tau }) , \theta) - T(\theta_\tau, \theta)  \|_2 . 
\end{eqnarray*} 
By \eqref{Thm2-eq-r02},  $ \sup_{\| \theta_{\tau}\|_1 \leq M ,  \| \theta  -\theta_0 \|_1 \leq \delta_n } \| T ( L (\theta_{\tau }) , \theta) \|_2$ is the maximum of finite number of $o_p(1)$. 
Thus,  we have
\begin{equation*}
 \sup_{\| \theta_{\tau}\|_1 \leq M ,  \| \theta  -\theta_0 \|_1 \leq \delta_n } \| T ( L (\theta_{\tau }) , \theta) \|_2  =o_p(1).
 \end{equation*}
By Assumption \ref{assumption-2},  we have
\begin{equation*}
	\sup_{\| \theta_{\tau}\|_1 \leq M ,  \| \theta  -\theta_0 \|_1 \leq \delta_n } \| G(L (\theta_{\tau }) , \theta )  - G(\theta_{\tau }, \theta ) \| _2 \leq C d +o_p(1).
\end{equation*}
We have
\begin{eqnarray*}
 &&  \| \hat{G} _{n,m} (L (\theta_{\tau }) , \theta )  - \hat{G} _{n,m} (\theta_{\tau }, \theta ) \| _2  \cr
 &&  \leq   \left \| \frac{1}{n- 1}  \sum_{i=2}^n \left  \{ \I (Y_i   < L(\theta_\tau) ^{\top} \hat{A}_{m,i}  (\theta, a_0) )  - \I (Y_i   < \theta_\tau^{\top} \hat{A}_{m,i}  (\theta, a_0) ) \right \} \hat{A}_{m,i} (\theta, a_0)  \right \| _2  \cr
 && \leq    \Big \| \frac{1}{n- 1}  \sum_{i=2}^n \left  \{ \I (Y_i   < L(\theta_\tau) ^{\top} \hat{A}_{m,i}  (\theta, a_0) )  - \I (Y_i   <  ( L(\theta_\tau)+ d M \mathbf{1}_4 )^{\top} \hat{A}_{m,i}  (\theta, a_0) ) \right \} \cr
 && \qquad \qquad \qquad \qquad \qquad \qquad \qquad \qquad  \qquad \qquad \times \hat{A}_{m,i} (\theta, a_0)  \Big\| _2  ,
\end{eqnarray*}
where the second inequality is due to $\I( Y_i < \cdot)$ is a monotone function and $\mathbf{1}_4$ is the $4 \times 1$ vector of all 1's. 
Thus, we have
\begin{eqnarray*}
 && \sup_{\| \theta_{\tau}\|_1 \leq M ,  \| \theta  -\theta_0 \|_1 \leq \delta_n } \| \hat{G} _{n,m} (L (\theta_{\tau }) , \theta )  - \hat{G} _{n,m} (\theta_{\tau }, \theta ) \| _2\cr
 &&   \leq  \sup_{\| \theta_{\tau}\|_1 \leq M ,  \| \theta  -\theta_0 \|_1 \leq \delta_n } \| T ( L (\theta_{\tau }) , \theta) -  T ( L (\theta_{\tau }) +d M \mathbf{1}_4 , \theta) \|_2   + O_p( d)   \cr
 && \leq O_p( d) + o_p(1), 
\end{eqnarray*}
and 
\begin{equation*}
 \sup_{\theta_{\tau} \in \Theta_{\tau}, \| \theta  -\theta_0 \|_1 \leq \delta_n }  \|\hat{G} _{n,m} (\theta_{\tau }, \theta,a_0  ) - G(\theta_{\tau }, \theta ) \|_2 =O_p( d)  + o_p(1).   
\end{equation*}
Since $d$ is arbitrarily small, we can show 
\begin{equation*} 
 \sup_{\theta_{\tau} \in \Theta_{\tau}, \| \theta  -\theta_0 \|_1 \leq \delta_n }  \|\hat{G} _{n,m} (\theta_{\tau }, \theta,a_0  ) - G(\theta_{\tau }, \theta ) \|_2 =  o_p(1).   
\end{equation*}
Since $G$ is a continuous function and $\theta_{\tau 0}$ is a unique solution, by Theorem 1 in \citep{chen2003estimation}, we can show the consistency of $\hat{\theta}_{\tau}$.

Now, we investigate the convergence rate of $\hat{\theta}_{\tau} $. 
We have
\begin{eqnarray*}
	G  (\hat{\theta}_{\tau}  ,\theta_0 )  &=& G (\theta_{\tau  0 }, \theta_0) +  \frac{\partial G  ( \theta_{\tau 0 }, \theta_0) }{\partial \theta_\tau} ( \hat{\theta}_{\tau   }- \theta_{\tau  0 } )  + O_p ( \| \hat{\theta}_{\tau   }- \theta_{\tau  0 }\| _2^2  ) \cr
	&=&\frac{\partial G  ( \theta_{\tau 0 }, \theta_0) }{\partial \theta_\tau} ( \hat{\theta}_{\tau   }- \theta_{\tau  0 } )  + O_p ( \| \hat{\theta}_{\tau   }- \theta_{\tau  0 }\| _2^2  ).
\end{eqnarray*}
Thus,  the convergence rate of $\hat{\theta}_{\tau}$ is the same as that of $G  (\hat{\theta}_{\tau   } ,\theta_0 ) $.
We have 
\begin{eqnarray*}
	\| G  (\hat{\theta}_{\tau   } ,\theta_0 ) \|_2  &\leq&  \|G  (\hat{\theta}_{\tau   } ,\theta_0 )  -G( \hat{\theta}_\tau, \hat{\theta})\|_2 + \|G( \hat{\theta}_\tau, \hat{\theta})   -G_n (\hat{\theta}_{\tau}, \hat{\theta}) + G_n (\theta_{\tau0}, \theta_0)\|_2 \cr
	&& \quad  + \| G_n (\theta_{\tau0}, \theta_0)\|_2 + \|G_n (\hat{\theta}_{\tau}, \hat{\theta}) \|_2 .
\end{eqnarray*}
First, consider $\|G  (\hat{\theta}_{\tau   } ,\theta_0 )  -G( \hat{\theta}_\tau, \hat{\theta})\|_2$.
By Taylor's expansion and Theorem \ref{Thm-1}, we have
\begin{eqnarray*}
 G  (\hat{\theta}_{\tau   } ,\theta_0 )  -G( \hat{\theta}_\tau, \hat{\theta})  &=& \frac{\partial G( \hat{\theta}_\tau, \theta_0)  }{\partial \theta ^{\top}} (\theta_0 - \hat{\theta} )  +O_p (n^{-1})  \cr
 &=&\frac{\partial G( \theta_{\tau 0} , \theta_0)  }{\partial \theta^{\top}} (\theta_0 - \hat{\theta} )  +O_p (n^{-1}  + n^{-1/2} \| \hat{\theta}_{\tau   }- \theta_{\tau  0 }\| _2 ).
\end{eqnarray*}
Thus, we have
\begin{equation}\label{Thm2-eq0001}
	\|G  (\hat{\theta}_{\tau   } ,\theta_0 )  -G( \hat{\theta}_\tau, \hat{\theta})\|_2 = O_p(n^{-1/2}). 
\end{equation}

Consider $ \| G_n (\theta_{\tau0}, \theta_0)\|_2$. 
We have
\begin{eqnarray*}
	-G_n (\theta_{\tau0}, \theta_0) = \frac{1}{n- 1}  \sum_{i=2}^n \left  \{   \I (Y_i   < \theta_{\tau 0}^{\top} A_i (\theta_0))  - \tau   \right \} A_i (\theta_0) .
\end{eqnarray*}
Thus, by the martingale convergence theorem, we have
\begin{equation}\label{Thm2-eq0002}
	\| G_n (\theta_{\tau0}, \theta_0) \|_2 = O_p(n^{-1/2}). 
\end{equation}

Consider $\|	G( \hat{\theta}_\tau, \hat{\theta})  -G_n (\hat{\theta}_{\tau}, \hat{\theta}) + G_n (\theta_{\tau0}, \theta_0)\|_2$. 
Similar to \eqref{Thm2-eq001}, we can show, for some $r >2$, 
\begin{eqnarray}  \label{Thm2-eq00001}
	&& E \[  \sup_{ (\theta_\tau ^{\prime}, \theta^{\prime}) :   \| \theta_{\tau} ^{\prime}  - \theta_{\tau}   \|_2  \leq \delta, \| \theta ^{\prime}  - \theta  \|_2  \leq \delta }  | m_j (A_i ,\theta_{\tau}^{\prime}, \theta ^{\prime} )- m_j (A_i ,\theta_{\tau} , \theta )|  ^r\] \cr
	&& \leq  C  E \[  \sup_{ (\theta_\tau ^{\prime}, \theta^{\prime}) :   \| \theta_{\tau} ^{\prime}  - \theta_{\tau}   \|_2  \leq \delta, \| \theta ^{\prime}  - \theta  \|_2  \leq \delta }   \left |    \I (Y_i   < \theta_{\tau   }^{ \prime \top} A_i (\theta ^{\prime}  ))  -  \I (Y_i   < \theta_{\tau  }^{\top} A_i  (\theta   ) )    \right |  \]  +C \delta^r   \cr
	&& \leq C     \sup_{ (\theta_\tau ^{\prime}, \theta^{\prime}) :   \| \theta_{\tau} ^{\prime}  - \theta_{\tau}   \|_2  \leq \delta, \| \theta ^{\prime}  - \theta  \|_2  \leq \delta }    E \[      F_{Y|X}  (  \theta_{\tau   }^{ \prime \top} A_i (\theta ^{\prime}  ))  -   F_{Y|X}( \theta_{\tau  }^{\top} A_i  (\theta   ) )    \]  + C \delta^r \cr
	&& \leq C    \delta.
\end{eqnarray}
Then, since $\( IV_i, OV_i, Y^2_i \)$ has exponentially decaying $\beta$-mixing,  by Lemma 4.2 \citep{chen2007large},  we have
\begin{eqnarray*} 
	   \sup_{\| \theta_\tau  - \theta_{\tau 0}\|_2  \leq \delta,  \| \theta  - \theta_0  \|_2  \leq \delta }  \frac{\sqrt{n} \| G(  \theta _\tau, \theta )  -G_n ( \theta _{\tau},  \theta ) + G_n (\theta_{\tau 0}, \theta_0) \|_2 } {1+ \sqrt{n} \{  \| G_n ( \theta _\tau,  \theta )\|_2 + \| G(\theta_{\tau},  \theta) \|_2  \} } =o_p(1). 
\end{eqnarray*}
Thus, we have
\begin{eqnarray}\label{Thm2-eq003}
	&&\|G( \hat{\theta}_\tau, \hat{\theta})   -G_n (\hat{\theta}_{\tau}, \hat{\theta}) + G_n (\theta_{\tau0}, \theta_0) \|_2  \cr
	 &&\leq o_p (1)  \{  \| G_n ( \hat{ \theta} _\tau,  \hat{\theta} )\|_2 + \| G(\hat{\theta}_{\tau},  \hat{\theta}) \|_2  \}  + o_p (n^{-1/2}) \cr
		&&=   o_p(n^{-1/2}  +\| \hat{\theta}_{\tau} -\theta_{\tau 0} \|_2) ,
\end{eqnarray}
where the last equality is due to \eqref{Thm2-eq003-1} and \eqref{Thm2-eq003-2} below. 
Similar to \eqref{Thm2-eq00001}, we can show
\begin{eqnarray}\label{Thm2-eq003-1}
 \| G_n ( \hat{ \theta} _\tau,  \hat{\theta} )\|_2 &\leq &  \| G_n (  \theta  _{\tau 0} , \theta_0 )\|_2 +  \| G_n ( \hat{ \theta} _\tau,  \hat{\theta} ) - G_n (  \theta  _{\tau 0} , \theta_0 ) \|_2   \cr
 	 &\leq&  \| G_n (  \theta  _{\tau 0} , \theta_0 )\|_2+ O_p (\| \hat{\theta}_{\tau} -\theta_{\tau 0} \|_2 + \| \hat{\theta} - \theta_0 \|_2) \cr
  &=&  O_p ( n^{-1/2}  +\| \hat{\theta}_{\tau} -\theta_{\tau 0} \|_2 )  , 
\end{eqnarray}
where the last equality is due to   Theorem \ref{Thm-1}.
By  Theorem \ref{Thm-1}, we have
\begin{eqnarray}\label{Thm2-eq003-2}
	\| G(\hat{\theta}_{\tau},  \hat{\theta}) \|_2  &\leq& \| G( \theta_{\tau 0 },   \theta_0) \|_2 + O_p (n^{-1/2} +\| \hat{\theta}_{\tau} -\theta_{\tau 0} \|_2)  \cr
	&=&  O_p (n^{-1/2} +\| \hat{\theta}_{\tau} -\theta_{\tau 0} \|_2). 
\end{eqnarray}

Finally, consider $ \|G_n (\hat{\theta}_{\tau}, \hat{\theta}) \|_2 $. 
Similar to proofs of \eqref{Thm2-eq-r01}, we can show
\begin{eqnarray*}
	 &&\|G_n (\hat{\theta}_{\tau}, \hat{\theta}) \|_2  \cr
	 &&\leq \|  G_n (\hat{\theta}_{\tau}, \hat{\theta})  - \hat{G}_{n,m} (\hat{\theta}_{\tau}, \hat{\theta}, a_0) \|_2  \cr
	 &&\leq \Big \|\frac{1}{n-1}  \sum_{i=2}^n \I \( |Y_i-  \theta_\tau ^{\top} A_{i} (\theta) | < C \| A_{i} (\theta) - \hat{A}_{m,i} (\theta, h_0)\|_1 + o(n^{-1/2})  \)   A_i (\theta) \Big \|_2 + o_p (n^{-1/2}) .
\end{eqnarray*}
Since $\sup_{\theta, i } \| A_{i} (\theta) - \hat{A}_{m,i} (\theta, h_0)\|_1  = O_p(m^{-1/2})$, it is enough to show the statement under $\| A_{i} (\theta) - \hat{A}_{m,i} (\theta, h_0)\|_1 \leq C m^{-1/2}$. 
Let
\begin{eqnarray*}
	&& G_n ^\prime (\theta_\tau, \theta)=  \frac{1}{n-1}  \sum_{i=2}^n  \{ \I \( |Y_i-  \theta_\tau ^{\top} A_{i} (\theta) | < C  ( m^{-1/2} + n^{-c})  \)    - \tau ^{\prime} \}   A_i (\theta) , \cr
	&& G^{\prime} (\theta_{\tau}, \theta) = E \Big [ \{F_{Y|X} ( \theta_\tau ^{\top} A_{i} (\theta) +C ( m^{-1/2} + n^{-c})   ) -   F_{Y|X} ( \theta_\tau ^{\top} A_{i} (\theta)  - C ( m^{-1/2} + n^{-c})  )  \cr
		&& \qquad \qquad \qquad \qquad \qquad \qquad \qquad \qquad \qquad \qquad   -\tau ^{\prime} \} A_i (\theta) \Big ], 
\end{eqnarray*}
where $c>1/2$ and $\tau ^{\prime} =   E \[ \I \( |Y_i-  \theta_{\tau_0} ^{\top} A_{i} (\theta_0) | < C  ( m^{-1/2} + n^{-c})  \) \] $.
Then, similar to the proofs of \eqref{Thm2-eq003}, we have
\begin{eqnarray*} 
	&&\|G ^{\prime} ( \hat{\theta}_\tau, \hat{\theta})   -G_n ^{\prime}(\hat{\theta}_{\tau}, \hat{\theta}) + G_n ^{\prime}(\theta_{\tau0}, \theta_0) \|_2  \cr
		&&=   o_p(n^{-1/2}  +\| \hat{\theta}_{\tau} -\theta_{\tau 0} \|_2) .
\end{eqnarray*}
Furthermore, we have
\begin{eqnarray*}
	&& E \[ G_n ^{\prime}(\theta_{\tau0}, \theta_0)  \] \leq C ( m^{-1/2} + n^{-c}), \cr
	&&G^{\prime} (\hat{\theta}_{\tau}, \hat{\theta})  = O (m^{-1/2} + n^{-c}) , \cr
	&& \tau^{\prime} = O ( m^{-1/2} + n^{-c}).
\end{eqnarray*}
Thus, we have
\begin{equation} \label{Thm2-eq00003}
	 \|G_n^{\prime} (\hat{\theta}_{\tau}, \hat{\theta}) \|_2 =o_p( n^{-1/2} + \| \hat{\theta}_{\tau} -\theta_{\tau 0} \|_2).
\end{equation}
Therefore, by \eqref{Thm2-eq0001}, \eqref{Thm2-eq0002}, \eqref{Thm2-eq003}, and \eqref{Thm2-eq00003},  we have
\begin{equation*}  
\| \hat{\theta}_{\tau} - \theta_{\tau 0} \|_2  =O_p ( n^{-1/2} ).
\end{equation*}

Now, we investigate the asymptotic normality. 
By \eqref{Thm2-eq0001}, \eqref{Thm2-eq0002}, \eqref{Thm2-eq003}, and \eqref{Thm2-eq00003}, we have
\begin{eqnarray*}
  &&\sqrt{n} ( \hat{\theta}_{\tau   }- \theta_{\tau  0 } ) \cr
  &&= \frac{\partial G( \theta_{\tau 0} , \theta_0)  }{\partial \theta_{\tau}^{\top}} ^{-1} \sqrt{n} G  (\hat{\theta}_{\tau   } ,\theta_0 ) + o_p(1) \cr
  	&&=  \frac{\partial G( \theta_{\tau 0} , \theta_0)  }{\partial \theta_{\tau}^{\top}} ^{-1} \[  \frac{1}{\sqrt{n}}  \sum_{i=2}^n \left  \{   \I (Y_i   < \theta_{\tau 0}^{\top} A_i (\theta_0))  - \tau   \right \} A_i (\theta_0) + \frac{\partial G( \theta_{\tau 0} , \theta_0)  }{\partial \theta^{\top}}  \sqrt{n} (\theta_0 - \hat{\theta} )  \] + o_p(1) \cr
  	&&=  \frac{1}{f_{\epsilon} ( F_{\epsilon} ^{-1}  (\tau) ) }  \Gamma_1 ^{-1} \frac{1}{\sqrt{n}}  \sum_{i=2}^n \left  \{  \tau -  \I (Y_i   < \theta_{\tau 0}^{\top} A_i (\theta_0))    \right \} A_i (\theta_0) -\Gamma_1 ^{-1} \Gamma_2 \sqrt{n} (\hat{\theta}  - \theta_0 ) + o_p(1).
\end{eqnarray*}
The statement \eqref{Thm-2-result1} is showed. 
The asymptotic normality is immediately showed by the martingale central limit theorem. 
\endpf

 \subsection{Proof of Theorem \ref{Thm-3}}

The proof of Theorem \ref{Thm-3} is the same as the proof of Theorem \ref{Thm-2}. 
Thus, we omit the proof.

\section*{Acknowledgments} 
 
The research of Donggyu Kim was supported in part by the National Research Foundation of Korea (NRF) (2021R1C1C1003216).

\section*{Data availability statement}
The intraday data is provided by the  Wharton Data Service (WRDS)
(web link: https://wrds-www.wharton.upenn.edu/). Please note that the data sharing policy of WRDS
restricts the redistribution of data.


\bibliography{myReferences}
\end{spacing}
\end{document}